\documentclass[twocolumn,aps,superscriptaddress,showpacs]{revtex4}
\usepackage{amsmath,bm}
\usepackage{graphicx}

\begin{document}

\title{Probing the Nuclear Symmetry Energy with Heavy-Ion Reactions Induced
by Neutron-Rich Nuclei}
\author{Lie-Wen Chen}
\affiliation{Institute of Theoretical Physics, Shanghai Jiao Tong University, Shanghai
200240, China}
\affiliation{Department of Physics, Texas A\&M University-Commerce, Commerce, Texas
75429-3011, USA}
\affiliation{Center of Theoretical Nuclear Physics, National Laboratory of Heavy Ion
Accelerator, Lanzhou 730000, China}
\author{Che Ming Ko}
\affiliation{Cyclotron Institute and Physics Department, Texas A\&M University, College
Station, Texas 77843-3366, USA}
\author{Bao-An Li}
\affiliation{Department of Physics, Texas A\&M University-Commerce, Commerce, Texas
75429-3011, USA}
\author{Gao-Chan Yong}
\affiliation{Institute of Modern Physics, Chinese Academy of Science, Lanzhou 730000,
China}
\affiliation{Graduate School, Chinese Academy of Science, Beijing 100039, P.R. China}
\affiliation{Department of Physics, Texas A\&M University-Commerce, Commerce, Texas
75429-3011, USA}
\date{\today}

\begin{abstract}
Heavy-ion reactions induced by neutron-rich nuclei provide a unique means to
investigate the equation of state of isospin-asymmetric nuclear matter,
especially the density dependence of the nuclear symmetry energy. In
particular, recent analyses of the isospin diffusion data in heavy-ion
reactions have already put a stringent constraint on the nuclear symmetry
energy around the nuclear matter saturation density. We review this exciting
result and discuss its implications on nuclear effective interactions and
the neutron skin thickness of heavy nuclei. In addition, we also review the
theoretical progress on probing the high density behaviors of the nuclear
symmetry energy in heavy-ion reactions induced by high energy radioactive
beams.
\end{abstract}

 \maketitle

\section{Introduction}

\label{intro}

During the last decade, many radioactive beam facilities have become
available in the world. At these facilities, nuclear reactions
involving nuclei with large neutron or proton excess can be studied,
providing the opportunities to study the properties of nuclear
matter under the extreme condition of large isospin asymmetry. This
has led to a lot of interests and activities in a new research
direction in nuclear physics, namely the isospin physics. The
ultimate goal of studying isospin physics is to extract information
on the isospin dependence of in-medium nuclear effective
interactions as well as the equation of state (EOS) of isospin
asymmetric nuclear matter, particularly its isospin-dependent term,
i.e., the density dependence of the nuclear symmetry energy. There
are already extensive reviews on the isospin physics in nuclear
physics, and they can be found in, e.g.,
Refs.\cite{LiBA98,LiBA01b,Dan02a,Lat04,Bar05,Ste05a}.

Knowledge about the nuclear symmetry energy extracted from the EOS
of isospin asymmetric nuclear matter is essential in understanding
not only many aspects of nuclear physics, such as heavy-ion
collisions induced by radioactive nuclei and the structure of exotic
nuclei, but also a number of important issues in astrophysics, such
as nucleosynthesis during pre-supernova evolution of massive stars
and the cooling of protoneutron stars. Although the nuclear symmetry
energy at normal nuclear matter density is known to be around $30$
MeV from the empirical liquid-drop mass formula \cite{Mey66,Pom03},
its values at other densities, especially at supra-normal densities,
are poorly known \cite{LiBA98,LiBA01b}. Predictions based on various
many-body theories differ widely at both low and high densities
\cite{Bom01,Die03}. Empirically, the incompressibility of asymmetric
nuclear matter is essentially undetermined \cite{Shl93}, even though
the incompressibility of symmetric nuclear matter at its saturation
density $\rho _{0}\approx 0.16$ fm$^{-3}$ has been determined to be
$231\pm 5 $ MeV from nuclear giant monopole resonances (GMR)
\cite{You99} and the EOS at densities of $2\rho _{0}<\rho <5\rho
_{0}$ has also been constrained
by measurements of collective flows in nucleus-nucleus collisions \cite%
{Dan02a}.

Theoretical studies of the EOS of isospin asymmetric nuclear matter
were started by Brueckner \textit{et al.} \cite{Bru67} and Siemens
\cite{Sie70} in the late 60's. Since then, There have been many
studies on this subject based on different many-body theories using
various two-body and three-body forces or interaction Lagrangians.
These many-body theories provide very useful tools for understanding
the properties of hot and dense nuclear matter, and they can be
roughly classified into three categories: the microscopic many-body
approach, the effective-field theory approach, and the
phenomenological approach. In the microscopic many-body approach,
the nuclear many-body problem is treated microscopically using
nucleon-nucleon interactions fitted to high-precision experimental
data and is thus free of parameters. The microscopic many-body
approach mainly includes the non-relativistic Brueckner-Hartree-Fock
(BHF) approach \cite{Sjo74,Cug87,Bom91,Zuo02},
relativistic Dirac-Brueckner-Hartree-Fock (DBHF) approach \cite%
{Mut87,Har87,Sum92,Hub93,Fuc04,Ma04,Sam05a}, self-consistent Green's
function approach \cite{Mut00,Dew02,Die03,Car03,Dic04}, and
variational many-body approach
\cite{Fri81,Lag81,Wir88a,Akm98,Muk07}. In the effective-field theory
approach, an effective interaction is constructed based on the
effective-field theory (EFT), leading to a systematic expansion of
the EOS in powers of density (the Fermi momentum). The
effective-field theory approach can be based on the density
functional theory \cite{Ser97,Fur04} or on chiral perturbation
theory \cite{Pra87,Lut00,Fin04,Vre04,Fri05,Fin06}. Since this
approach can be linked to low energy QCD and its symmetry breaking,
it has the advantage of small number of free parameters and a
correspondingly higher predictive power. The phenomenological
approach is based on effective density-dependent nuclear forces or
effective interaction Lagrangians. In these approaches, a number of
parameters have to be adjusted to fit the properties of many nuclei.
This type of models mainly includes the relativistic mean-field
(RMF) theory
\cite{Ser86,Sum92,Chi77,Hor87,Gle82,Hir91,Sug94,Rei89,Rin96},
relativistic and non-relativistic Hartree-Fock \cite%
{Mil74,Bro78,Jam81,Hor83,Bou87,Lop88,Ber93,Wer94,Kho96,Vau72,Bra85,Sto07}
or Thomas-Fermi approximations \cite{Bra85,Kol85,Ban90}. These
phenomenological approaches allow the most precise description for
the properties of finite nuclei. Both the phenomenological and EFT
approaches contain parameters that are fixed by nuclear properties
around the saturation density and thus usually give excellent
descriptions for the nuclear properties around or below the
saturation density. Their predictions at the high density region
are, however, probably unreliable. In addition, due to different
approximations or techniques used in different microscopic many-body
approaches, their predictions on the properties of nuclear matter
could be very different even for the same bare nucleon-nucleon
interaction \cite{Die03,LiZH06}. In particular, predictions on the
properties of isospin asymmetric nuclear matter, especially the
density dependence of the nuclear symmetry energy, are still
significantly different for different many-body theory approaches.

Fortunately, heavy-ion reactions induced by radioactive beams
provide a unique opportunity to investigate in terrestrial
laboratories the EOS of asymmetric nuclear matter, particularly the
density dependence of the nuclear symmetry energy. During the past
decade, a large amount of theoretical and experimental efforts have
been devoted to the study of the properties of isospin asymmetric
nuclear matter
via heavy-ion reactions \cite{LiBA95,LiBA96,LiBA97a,LiBA97b,Che97,LiBA98,Che98,Bar98,Che99a,Zha99,Che00,Zha00,LiBA01b,Bar05,Mul95,Xu00,Tan01a,Bar02,Tsa01,LiBA01a,LiBA00,Tan01b,LiBA02,Che03a,Che03b,Ono03,Liu03,Che04,LiBA04a,Shi03,LiBA04b,Riz04,LiBA05a,LiBA05b,Zha05,LiQF05a,Tia05,Gai04,LiQF05b,Fer05,LiBA05e,Yon07}%
. To extract information about the EOS of neutron-rich matter,
especially the density dependence of the nuclear symmetry energy,
from heavy-ion reactions induced by radioactive beams, one needs
reliable theoretical tools. Transport models that include explicit
isospin-dependent degrees of freedom are especially useful for
understanding the role of isospin degree of freedom in the dynamics
of central nuclear reactions induced by neutron nuclei at
intermediate and high energies and in extracting information about
the EOS of produced neutron-rich matter. During past two decades,
significant progresses have been made in developing semi-classical
transport models for nuclear reactions. These semi-classical models
mainly include the following two types: the
Boltzmann-Uehling-Ulenbeck (BUU) model \cite{Ber88b} and the quantum
molecular dynamical (QMD) model \cite{Aic91}. While it is important
to develop practically implementable quantum transport theories,
applications of the semi-classical transport models have enabled us
to learn a great deal of interesting physics from heavy-ion
reactions. In particular, with the development of the radioactive
nuclear beam physics, some isospin-dependent transport models
\cite{LiBA95,LiBA96,LiBA97a,Che98,Bar98,Riz04,LiBA05c,Zha05} have
been successfully developed in recent years to describe the nuclear
reactions induced by neutron nuclei at intermediate and high
energies.

In studying the properties of asymmetric nuclear matter from
heavy-ion reactions induced by neutron-rich nuclei, a key task is
to identify experimental observables that are sensitive to the
density dependence of the nuclear symmetry energy, especially at
high densities. Because of the fact that the symmetry potentials
for neutrons and protons have opposite signs and that they are
generally weaker than the nuclear isoscalar potential at same
density, most observables proposed so far use differences or
ratios of isospin multiplets of baryons, mirror nuclei and
mesons, such as the neutron/proton ratio of nucleon emissions \cite{LiBA97a}%
, neutron-proton differential flow \cite{LiBA00}, neutron-proton
correlation function \cite{Che03a}, $t$/$^{3}$He
\cite{Che03b,Zha05}, $\pi ^{-}/\pi ^{+}$
\cite{LiBA02,Gai04,LiBA05a,LiQF05b}, $\Sigma ^{-}/\Sigma ^{+}$
\cite{LiQF05a} and $K^{0}/K^{+}$ ratios \cite{Fer05}, etc.. In
addition, in order to reduce the systematical errors, multiple
probes taken from several reaction systems using different isotopes
of the same element have also been proposed. These multiple probes
mainly include double ratio or double differential flow.

Indeed, recent experimental and theoretical analysis of the isospin
diffusion data from heavy-ion reactions has led to significant
progress in determining the nuclear symmetry energy at subnormal
densities \cite{Tsa04,Che05a,LiBA05c}. Based on the same underlying
Skyrme interactions as the ones constrained by the isospin diffusion
data, the neutron-skin thickness in $^{208}$Pb calculated within the
Hartree-Fock approach is consistent with available experimental data
\cite{Ste05b,LiBA06a,Che05b}. This symmetry energy is also
consistent with that from a relativistic mean-field model using an
accurately calibrated parameter set that reproduces both the giant
monopole resonance in $^{90}$Zr and $^{208}$Pb, and the isovector
giant dipole resonance of $^{208}$Pb \cite{Tod05}. It further agrees
with the symmetry energy recently obtained from isoscaling analyses
of isotope ratios in intermediate-energy heavy ion collisions
\cite{She07}. These different studies have provided so far the best
phenomenological constraints on the symmetry energy at sub-normal
densities. Information on the symmetry energy at supra-normal
densities, on the other hand, remains inclusive and more efforts are
needed to investigate the supra-normal density behavior of the
symmetry energy. Heavy-ion collisions induced by future high energy
radioactive beams to be available at high energy radioactive beam
facilities will provide a unique opportunity for determining the
symmetry energy at supra-normal densities.

In the present paper, we review recent progress on the determination
of the nuclear symmetry energy in heavy-ion reactions induced by
neutron-rich nuclei. In particular, we review the exciting results
on density dependence of the nuclear symmetry energy at subnormal
densities determined from recent analysis of the isospin diffusion
data in heavy-ion reactions.  We also discuss the implications
derived from this new information on nuclear effective interactions
and the neutron skin thickness of heavy nuclei. In addition, we
review theoretical progress in studying the behavior of nuclear
symmetry energy at high density from heavy-ion reactions induced by
high energy radioactive beams.

The paper is organized as follows. In Section \ref{Esym}, we give a brief
introduction to the nuclear symmetry energy. We then describe in Section \ref%
{ibuu04} the IBUU04 hadron transport model for nuclear reactions
induced by radioactive beams at intermediate energies. In Section
\ref{diffusion}, we present the results from the IBUU04 model
analysis of the isospin diffusion data in heavy-ion reactions and
discuss the stringent constraint they have imposed on the nuclear
symmetry energy around the nuclear matter saturation density. Based
on the constrained symmetry energy from the isospin diffusion data,
we discuss in Section \ref{skyrmenskin} the implications of the
isospin diffusion data on nuclear effective interactions and the
neutron skin thickness of heavy nuclei. In Section
\ref{highdensity}, we review theoretical progress on studying the
behavior of the nuclear symmetry energy at high density in heavy-ion
reactions induced by high energy radioactive beams. Finally, a
summary is given in Section \ref{summary}.

\section{The nuclear symmetry energy}

\label{Esym}

In the parabolic approximation that has been verified by all
many-body theories to date, the EOS of isospin asymmetric nuclear
matter can be written as
\begin{equation}
E(\rho ,\delta )=E(\rho ,\delta =0)+E_{\text{sym}}(\rho )\delta
^{2}+O(\delta ^{4}),  \label{EsymPara}
\end{equation}%
where $\rho =\rho _{n}+\rho _{p}$ is the baryon density with $\rho
_{n}$ and $\rho _{p}$ denoting the neutron and proton densities,
respectively; $\delta =(\rho _{n}-\rho _{p})/(\rho _{p}+\rho _{n})$
is the isospin asymmetry; $E(\rho ,\delta =0)$ is the energy per
nucleon in symmetric nuclear matter, and
\begin{equation}
E_{\text{sym}}(\rho )=\frac{1}{2}\frac{\partial ^{2}E(\rho ,\delta )}{%
\partial \delta ^{2}}|_{\delta =0}  \label{sym}
\end{equation}%
is the nuclear symmetry energy. In Eq. (\ref{EsymPara}), there are
no odd-order $\delta $ terms due to the exchange symmetry of the
proton and neutron in the nuclear matter (the charge symmetry of
nuclear forces). Higher-order terms in $\delta $ are negligible. For
example, the magnitude of the $\delta ^{4}$ term at $\rho _{0}$ has
been estimated to be less than $1$ MeV \cite{Sie70,Sjo74,Lag81}. As
a good approximation, the density dependent symmetry energy
$E_{\text{sym}}(\rho )$ can thus be extracted from
$E_{\text{sym}}(\rho )\approx E(\rho ,\delta
=1)-E(\rho ,\delta =0)$ which implies that the symmetry energy $E_{\text{sym}%
}(\rho )$ is the energy cost to convert all protons in a symmetric
nuclear matter to neutrons at the fixed density $\rho $.

Around the nuclear matter saturation density $\rho _{0}$, the nuclear
symmetry energy $E_{\text{sym}}(\rho )$\ can be further expanded to
second-order as
\begin{equation}
E_{\text{sym}}(\rho )=E_{\text{sym}}(\rho _{0})+\frac{L}{3}\left( \frac{\rho
-\rho _{0}}{\rho _{0}}\right) +\frac{K_{\text{sym}}}{18}\left( \frac{\rho
-\rho _{0}}{\rho _{0}}\right) ^{2},  \label{EsymLK}
\end{equation}%
where $L$ and $K_{\text{sym}}$ are the slope and curvature
parameters of the nuclear symmetry energy at $\rho _{0}$, i.e.,
\begin{eqnarray}
L &=&3\rho _{0}\frac{\partial E_{\text{sym}}(\rho )}{\partial \rho }|_{\rho
=\rho _{0}},  \label{L} \\
K_{\text{sym}} &=&9\rho _{0}^{2}\frac{\partial ^{2}E_{\text{sym}}(\rho )}{%
\partial ^{2}\rho }|_{\rho =\rho _{0}}.  \label{Ksym}
\end{eqnarray}%
The $L$ and $K_{\text{sym}}$ characterize the density dependence of the
nuclear symmetry energy around normal nuclear matter density, and thus
provide important information on properties of the nuclear symmetry energy
at both high and low densities.

At the nuclear matter saturation density $\rho _{0}$ and around
$\delta =0$, the isobaric incompressibility of asymmetric nuclear
matter can be further expressed as \cite{Pra85,Lop88}
\begin{equation}
K(\delta )=K_{0}+K_{\text{asy}}\delta ^{2}
\end{equation}%
where $K_{0}$ is the incompressibility of symmetric nuclear matter
at the nuclear matter saturation density $\rho _{0}$ and the
isospin-dependent part $K_{\mathrm{asy}}\approx K_{\mathrm{sym}}-6L
$ \cite{Bar02} characterizes the density dependence of the nuclear
symmetry energy. In principle, the information on $K_{\mathrm{asy}}$
can be extracted experimentally by measuring the GMR of neutron-rich
nuclei and a constraint of $-566\pm 1350<K_{\mathrm{asy}}<139\pm
1617$ MeV has been extracted earlier from a systematic study on the
GMR of finite nuclei depending on the mass region of nuclei and the
number of parameters used in parameterizing the incompressibility of
finite nuclei \cite{Shl93}. The large uncertainties in the extracted
$K_{\mathrm{asy}}$ thus does not allow us to distinguish the
different nuclear symmetry energies in theoretical models.

\begin{figure}[tbh]
\includegraphics[scale=1.05]{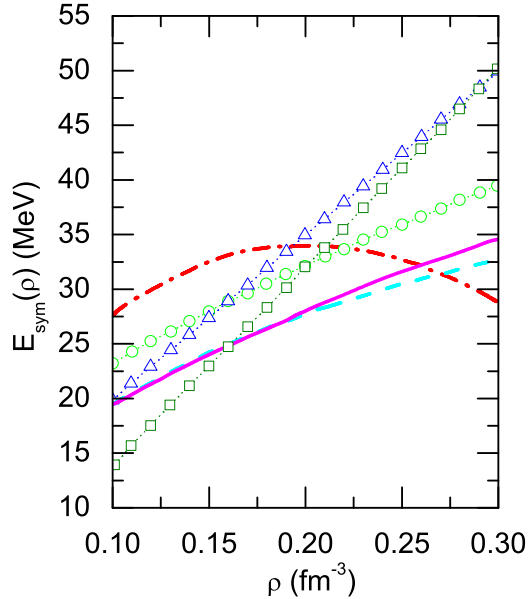} 
\caption{(Color online) Density dependence of the symmetry energy
from the continuous choice Brueckner-Hartree-Fock with Reid93
potential (circles), self-consistent Green's function theory with
Reid93 potential (full line), variational calculation with Argonne
Av14 potential (dashed line), Dirac-Brueckner-Hartree-Fock
calculation (triangles), relativistic mean-field model (squares),
and effective field theory (dash-dotted line). Data are taken from
\protect\cite{Die03}.} \label{piekEsym}
\end{figure}

Information about the nuclear symmetry energy is important not only
for nuclear physics, but also for a number of important issues in
astrophysics. For example, the prompt shock invoked to understand
the explosion mechanism of a type II supernova requires a relatively
soft EOS \cite{Bar85}, which can be understood in terms of the
dependence of the nuclear incompressibility on isospin. In the model
for prompt explosion \cite{Kah89}, the electron-capture reaction
drives the star in the latest stage of collapse to an equilibrium
state where the proton concentration is about $1/3 $, which,
according to microscopic many-body calculations, reduces the nuclear
matter incompressibility by about $30\%$ compared to that for
symmetric nuclear matter. Moreover, the magnitude of proton
concentration at $\beta $ equilibrium in a neutron star is almost
entirely determined by the symmetry energy. The proton fraction
affects not only the stiffness of the EOS but also the cooling
mechanisms of neutron stars \cite{Lat91,Sum94} and the possibility
of kaon condensation ($e^{-}\rightarrow K^{-}\nu _{e}$) in dense
stellar matter \cite{Lee96}. If the proton concentration is larger
than a critical value of about $15\%$, the direct URCA process
$(n\rightarrow p+e^{-}+\bar{\nu}_{e},~p+e^{-}\rightarrow n+\nu
_{e})$ becomes possible, and would then enhance the emission of
neutrinos, making it a more important process in the cooling of a
neutron star \cite{Lat91}.

Unfortunately, the density dependence of $E_{\mathrm{sym}}(\rho
)$, especially its high density behavior, is poorly known and is
regarded as the most uncertain among all properties of an isospin
asymmetric nuclear matter. Even around the saturation density,
values of the parameters $L$, $K_{\text{sym}}$, and
$K_{\mathrm{asy}}$ are still very uncertain with different
theoretical models giving very different predictions. This can be
seen in Fig.~\ref{piekEsym} where we show the density dependence
of the nuclear symmetry energy from some of the most widely used
microscopic many-body theories \cite{Die03}. One sees that the
theoretical predictions diverge widely at both low and high
densities. In fact, even the sign of the symmetry energy above
$3\rho _{0}$ is uncertain \cite{Bom01}. The theoretical
uncertainties are largely due to a lack of knowledge about the
isospin dependence of nuclear effective interactions and the
limitations in the techniques in solving the nuclear many-body
problem.

As mentioned in the Introduction, heavy-ion reactions, especially
those induced by radioactive beams, provide a unique opportunity to
pin down the density dependence of nuclear symmetry energy in
terrestrial laboratories. Indeed, significant progress has been made
recently both experimentally and theoretically in determining the
symmetry energy at subnormal densities. At sub-normal densities, a
density-dependent symmetry energy of $E_{\text{sym}}(\rho )\approx
31.6(\rho /\rho _{0})^{0.69}$ has been found to best reproduce both
the isospin diffusion \cite{Che05a,Tsa04,Ste05b,LiBA05c,Che05b} and
isoscaling \cite{She07} data in heavy-ion collisions as well as the
presently acceptable neutron-skin thickness in $^{208}$Pb
\cite{Ste05b,LiBA06a,Che05b}. Together, these results represent the
best phenomenological constraints available so far on the symmetry
energy at sub-normal densities. Although the high density behavior
of the symmetry energy remain largely undetermined, future high
energy radioactive beams to be available at high energy radioactive
beam facilities will allow us to determine the symmetry energy at
supra-normal densities.

\section{IBUU Transport Model for Nuclear Reactions Induced by Radioactive
Beams}

\label{ibuu04}

Transport models are useful theoretical tools not only for studying
the reaction mechanisms but also for extracting information on the
properties of produced hot dense matter in heavy ion collisions. For
nuclear reactions induced by radioactive beams, comparing
experimental data with transport model calculations allows us to
extract the information about the EOS of neutron-rich matter. The
IBUU model, which has been very useful in understanding a number of
new phenomena associated with the isospin degree of freedom in
heavy-ion reactions, is an isospin- and momentum-dependent transport
model that is based on the Boltzmann-Uhling-Uhlenbeck equation and
is applicable for heavy-ion reactions induced by both stable and
radioactive beams \cite{LiBA04a}.

In the IBUU model, besides nucleons, $\Delta $ and $N^{\ast }$
resonances as well as pions and their isospin-dependent dynamics are
included. The initial neutron and proton density distributions of
projectile and target nuclei are obtained from the Relativistic
Mean-field model or the Skyrme-Hartree-Fock model. It has the option
of using either the experimental free-space nucleon-nucleon (NN)
scattering cross sections or the in-medium NN cross sections. For NN
inelastic collisions, the experimental free-space cross sections are
used as their in-medium cross sections are still very much
controversial. The total and differential cross sections for all
other particles are taken either from experimental data or obtained
by using the detailed-balance formula. Time dependence of the
isospin-dependent phase-space distribution functions of involved
particles are solved numerically using the test-particle method. In
treating NN scattering, the isospin dependent Pauli blocking factors
for fermions is also included.

In the following, we outline the two major ingredients, i.e., the
single-nucleon potential and the NN cross sections, of the version
IBUU04 of the isospin- and momentum-dependent IBUU transport model
for nuclear reactions induced by radioactive beams \cite{LiBA04a}.
Other details, such as the initialization of the phase space
distributions of colliding nuclei, the Pauli blocking, {\it etc.}
can be found in Refs. \cite{LiBA97a,LiBA98,LiBA01b,LiBA04a,LiBA05c}.

\subsection{Single-nucleon potential}

\begin{figure}[tbh]
\includegraphics[scale=0.85]{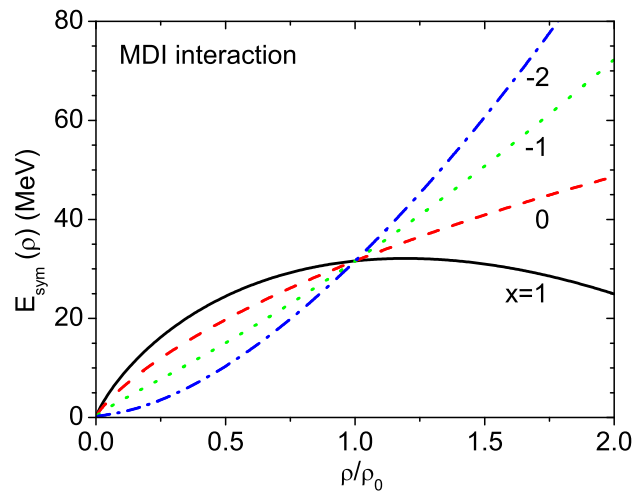}
\caption{(Color online) Symmetry energy as a function of density for the MDI
interaction with $x=1,0,-1$ and $-2$. Taken from Ref. \protect\cite{Che05a}.}
\label{MDIEsym}
\end{figure}

One of the most important inputs to all transport models is the
single-nucleon potential. Both the isovector (symmetry potential)
and isoscalar parts of this potential should be momentum-dependent
due to the non-locality of strong interactions and the Pauli
exchange effects in many-fermion systems. In the IBUU04, we use a
single-nucleon potential derived from the Hartree-Fock approximation
based on a modified Gogny effective interaction (MDI) \cite{Das03},
i.e.,
\begin{eqnarray}
U(\rho ,\delta ,\vec{p},\tau ,x) &=&A_{u}(x)\frac{\rho _{\tau ^{\prime }}}{%
\rho _{0}}+A_{l}(x)\frac{\rho _{\tau }}{\rho _{0}}  \notag  \label{mdi} \\
&+&B(\frac{\rho }{\rho _{0}})^{\sigma }(1-x\delta ^{2})-8\tau x\frac{B}{%
\sigma +1}\frac{\rho ^{\sigma -1}}{\rho _{0}^{\sigma }}\delta \rho _{\tau
^{\prime }}  \notag \\
&+&\frac{2C_{\tau ,\tau }}{\rho _{0}}\int d^{3}p^{\prime }\frac{f_{\tau }(%
\vec{r},\vec{p}^{\prime })}{1+(\vec{p}-\vec{p}^{\prime })^{2}/\Lambda ^{2}}
\notag \\
&+&\frac{2C_{\tau ,\tau ^{\prime }}}{\rho _{0}}\int d^{3}p^{\prime }\frac{%
f_{\tau ^{\prime }}(\vec{r},\vec{p}^{\prime })}{1+(\vec{p}-\vec{p}^{\prime
})^{2}/\Lambda ^{2}}.  \label{MDI}
\end{eqnarray}%
In the above $\tau =1/2$ ($-1/2$) for neutrons (protons) and $\tau
\neq \tau ^{\prime }$; $\sigma =4/3$; $f_{\tau }(\vec{r},\vec{p})$
is the phase space distribution function at coordinate $\vec{r}$
and momentum $\vec{p}$. The parameters
$A_{u}(x),A_{l}(x),B,C_{\tau ,\tau },C_{\tau ,\tau ^{\prime }}$
and $\Lambda $ were obtained by fitting the momentum dependence of
$U(\rho ,\delta ,\vec{p},\tau ,x)$ to that predicted by the Gogny
Hartree-Fock and/or the Brueckner-Hartree-Fock calculations, the
saturation properties of symmetric nuclear matter, and the
symmetry energy of 31.6 MeV at normal nuclear matter density $\rho
_{0}=0.16$ fm$^{-3}$\cite{Das03}. The incompressibility $K_{0}$ of
symmetric nuclear matter at $\rho _{0}$ is set to be 211 MeV. The
parameters $A_{u}(x)$ and $A_{l}(x)$, given by
\begin{equation}
A_{u}(x)=-95.98-x\frac{2B}{\sigma
+1},~~~~A_{l}(x)=-120.57+x\frac{2B}{\sigma +1},
\end{equation}
depend on the parameter $x$ that can be adjusted to mimic the
predicted $E_{\rm sym}(\rho ) $ from microscopic and/or
phenomenological many-body theories. The last two terms in Eq.
(\ref{MDI}) contain the momentum-dependence of the single-particle
potential. The momentum dependence of the symmetry potential stems
from the different interaction strength parameters $C_{\tau ,\tau
^{\prime }}$ and $C_{\tau ,\tau }$ for a nucleon of isospin $\tau $
interacting, respectively, with unlike and like nucleons in the
background fields. More specifically, we use $C_{\tau ,\tau ^{\prime
}}=-103.4$ MeV and $C_{\tau ,\tau }=-11.7$ MeV. With these
parameters, the isoscalar potential estimated from $(U_{\rm
neutron}+U_{\rm proton})/2$ agrees reasonably well with predictions
from the variational many-body theory \cite{Wir88b}, the more
advanced BHF approach \cite{Zuo05} including three-body forces, and
the Dirac-Brueckner-Hartree-Fock (DBHF) calculations\cite{Sam05a} in
a broad range of density and momentum. As an example, we show in
Fig.~\ref{MDIEsym} the density dependence of the symmetry energy for
$x=-2 $, $-1$, $0$ and $1$. It is seen that the symmetry energy
becomes softer with increasing value of the parameter $x$.

\begin{table}[tbp]
\caption{{\protect\small The parameters }$F${\protect\small \ (MeV), }$G$
{\protect\small , }$K_{\text{sym}}${\protect\small \ (MeV), } $L$%
{\protect\small \ (MeV), and }$K_{\text{asy}}${\protect\small \ (MeV) for
different values of} $x${\protect\small . Taken from Ref. \protect\cite%
{Che05a}.}}
\label{MDIx}%
\begin{tabular}{ccccccc}
\hline\hline
$x$ & \quad $F$ & $G$ & $K_{\text{sym}}$ & $L$ & $K_{\text{asy}}$ &  \\
\hline
$1$ & $107.232$ & $1.246$ & $-270.4$ & $16.4$ & -368.8 &  \\
$0$ & $129.981$ & $1.059$ & $-88.6$ & $62.1$ & -461.2 &  \\
$-1$ & $3.673$ & $1.569$ & $94.1$ & $107.4$ & -550.3 &  \\
$-2$ & $-38.395$ & $1.416$ & $276.3$ & $153.0$ & -641.7 &  \\ \hline\hline
\end{tabular}%
\end{table}

The interaction part of nuclear symmetry energy can be parameterized by
\begin{equation}
E_{\text{sym}}^{\mathrm{pot}}(\rho )=F(x)\rho /\rho _{0}+(18.6-F(x))(\rho
/\rho _{0})^{G(x)}  \label{esympot}
\end{equation}%
with $F(x)$ and $G(x)$ given in Table \ref{MDIx} for $x=1$, $0$, $-1$ and $%
-2 $. Also shown in Table \ref{MDIx} are other properties of the
symmetry energy, including its slope parameter $L$ and curvature
parameter $K_{\text{sym}}$ at $\rho _{0}$, as well as the
isospin-dependent part $K_{\mathrm{asy}}$ of the isobaric
incompressibility of asymmetric nuclear matter. It is seen that
the stiffness of the symmetry energy increases with decreasing $x$
values.

\begin{figure*}[tbh]
\includegraphics[scale=1.5]{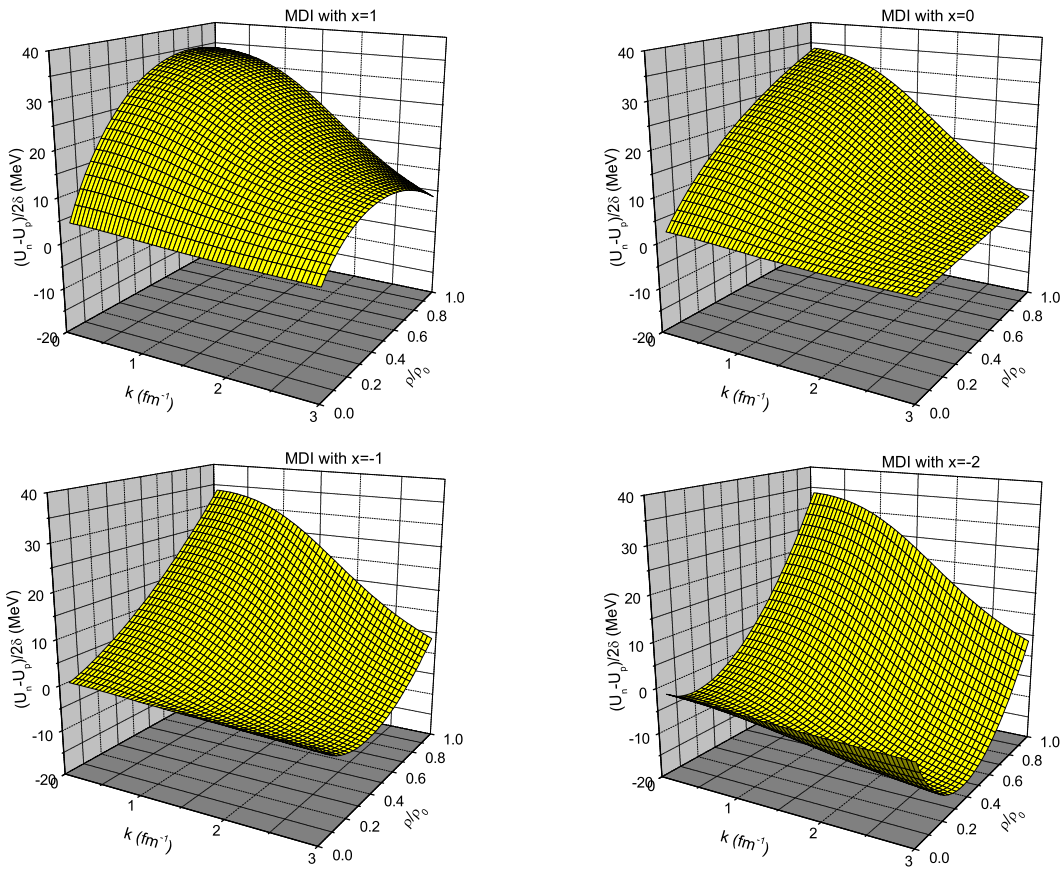}
\caption{(Color online) Symmetry potential as a function of momentum and
density for MDI interactions with $x=1,0,-1$ and $-2$. Taken from Ref.
\protect\cite{LiBA05c}.}
\label{SymPotRhoK}
\end{figure*}

\begin{figure}[tbh]
\includegraphics[scale=0.85]{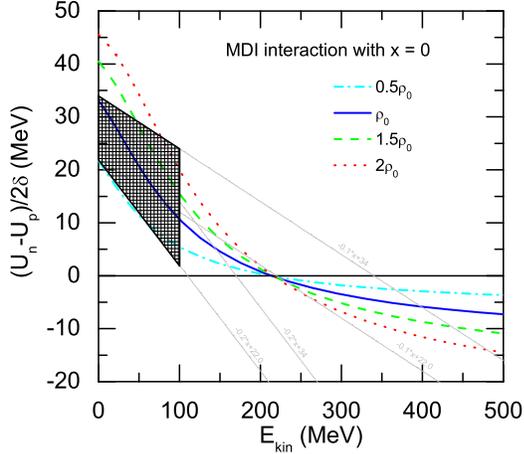}
\caption{(Color online) Kinetic energy dependence of $(U_{\rm
neutron}-U_{\rm proton})/2\protect\delta $ at different densities
using the MDI interaction with $x=0$. The shaded region indicates
the experimental constraint.} \label{LanePotEkin}
\end{figure}

What is particularly interesting and important for nuclear reactions induced
by neutron-rich nuclei is the isovector (symmetry) potential. The strength
of this potential can be estimated very accurately from $%
(U_{\rm neutron}-U_{\rm proton})/2\delta $ \cite{LiBA04a}. In
Fig.~\ref{SymPotRhoK}, the strength of the symmetry potential for
four $x$ parameters is displayed as a function of momentum and
density. Here we have only plotted the symmetry potential at
sub-saturation densities most relevant to heavy-ion reactions
studies at intermediate energies. The momentum dependence of the
symmetry potential is seen to be the same for all values of $x$ as
the parameter $x$ appears by construction only in the
density-dependent part of the single-nucleon potential given by
Eq.(\ref{mdi}). Systematic analysis of a large number of
nucleon-nucleus scattering experiments and (p,n) charge exchange
reactions at beam energies below about 100 MeV has shown that the
data can be well described by the parametrization $U_{\rm
Lane}=a-bE_{\rm kin}$ with $a\simeq 22-34$ MeV and $b\simeq
0.1-0.2$ \cite{Sat69,Hof72,Hod94,Kon03}. Although the
uncertainties in both parameters $a$ and $b$ are large, the
symmetry potential at $\rho _{0}$, i.e., the Lane potential,
clearly decreases approximately linearly with increasing beam
energy $E_{\rm kin}$. This provides a stringent constraint on the
symmetry potential. The potential in Eq. (\ref{mdi}) at $\rho
_{0}$ satisfies this requirement very well as seen in
Fig.~\ref{SymPotRhoK}. This can be more clearly seen from the
solid line in Fig.~\ref{LanePotEkin} which gives the kinetic
energy dependence of $(U_{\rm neutron}-U_{\rm proton})/2\delta $
at normal nuclear saturation density given by the MDI interaction
with $x=0$.  Also shown in this figure are the predicted kinetic
energy dependence of the symmetry potential from the MDI
interaction with $x=0$ for densities away from normal nuclear
density, which are presently not known empirically. Experimental
determination of both the density and momentum dependence of the
symmetry potential is thus of great interest, and heavy-ion
reactions with radioactive beams provides a unique tool to explore
this information in terrestrial laboratories.

Although the effective mass of a nucleon in nuclear matter depends
the density of nuclear matter as well as the momentum of the nucleon
\cite{Jam89,Neg98,Fuc04}, the momentum dependence of the symmetry
potential further leads to different effective masses for neutrons
and protons in isospin asymmetric nuclear matter, i.e.,
\begin{equation}
\frac{m_{\tau }^{\ast }}{m_{\tau }}=\left\{ 1+\frac{m_{\tau }}{p}\frac{%
dU_{\tau }}{dp}\right\} .  \label{emass}
\end{equation}%
When the effective mass is evaluated at the Fermi momentum $p_{\tau
}=p_{\rm F}({\tau })$, Eq. (\ref{emass}) yields the Landau mass
which is related to the $f_{1}$ Landau parameter of a Fermi liquid
\cite{Jam89,Neg98,Fuc04}. A detailed discussion about different
kinds of effective masses can be found in Ref. \cite{Jam89}. With
the potential in Eq. (\ref{mdi}), the nucleon effective masses are
independent of the $x$ parameter as the momentum-dependent part of
the nuclear potential is independent of the parameter $x$.

\begin{figure}[tbh]
\includegraphics[height=0.35\textheight,angle=-90]{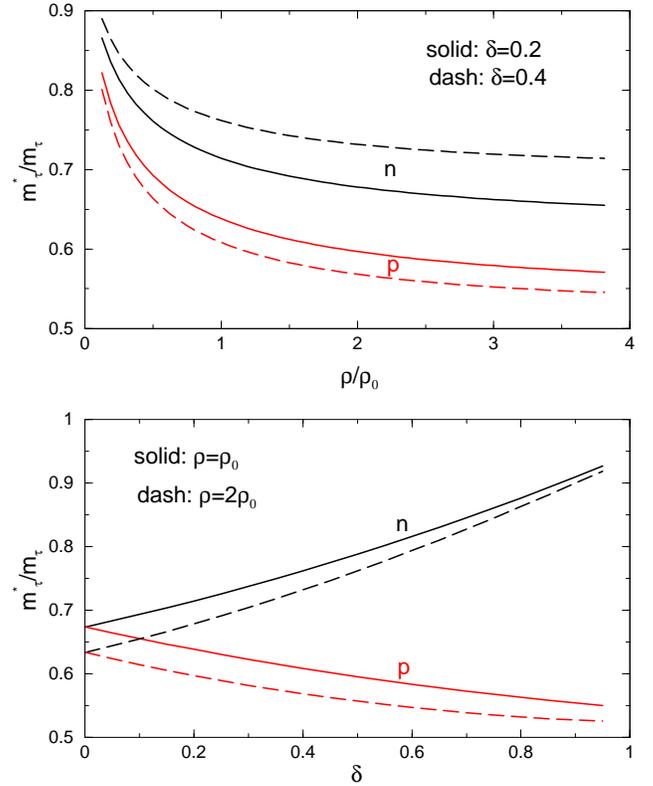}
\caption{(Color online) Neutron and proton effective masses in
asymmetric matter as a function of density (upper window) and
isospin asymmetry (lower window). Taken from Ref.
\protect\cite{LiBA05c}.} \label{EffMass}
\end{figure}

Shown in Fig.~\ref{EffMass} are the effective masses of neutrons and
protons at their respective Fermi surfaces as functions of density
(upper window) and isospin asymmetry (lower window) obtained from
the MDI interaction.  It is seen that the effective mass of neutrons
is larger than that of protons and the splitting between them
increases with both the density and isospin asymmetry of the medium.
Although the momentum dependence of the symmetry potential and the
associated splitting between the neutron and proton effective masses
are still highly controversial among different approaches and/or
using different nuclear effective interactions
\cite{Riz04,LiBA04c,Beh05}, the results presented here are
consistent with the predictions from all non-relativistic
microscopic models, see, e.g., \cite{Bom91,Zuo05,Sjo76}, and the
non-relativistic limit of microscopic relativistic many-body
theories, see, e.g., \cite{Sam05a,Ma04,Fuc04}. Recent transport
model studies have indicated that the neutron/proton ratio at high
transverse momenta and/or rapidities is a potentially useful probe
of the splitting between the neutron and proton effective masses in
neutron-rich matter \cite{LiBA04a,Bar05}.

The effect due to the momentum dependence of nuclear mean-field
potential can be studied by comparing its predictions with those
obtained using the momentum-independent nucleon potential $U(\rho
,\delta ,\tau )\equiv U_{0}(\rho )+U_{\text{sym}}^{\text{MDI}
(x)}(\rho ,\delta ,\tau )$ with the isoscalar part $U_{0}(\rho )$
taken from the original momentum-independent soft nuclear
potential with $K_{0}=200$ MeV (SBKD) introduced by Bertsch, Kruse
and Das Gupta \cite{Ber84}, i.e.,
\begin{equation}
U_{\text{SBKD}}(\rho )=-356\rho /\rho _{0}+303(\rho /\rho _{0})^{7/6}.
\end{equation}%
For the momentum-independent symmetry potential $U_{\text{sym}}^{\text{MDI}%
(x)}(\rho ,\delta ,\tau )$, it can be obtained from $U_{\text{sym}}^{\text{%
MDI}(x)}(\rho ,\delta ,\tau )=\partial W_{\text{sym}}/\partial \rho _{\tau }$
using the isospin-dependent part of the potential energy density $W_{\text{%
sym}}=E_{\text{sym}}^{pot}(\rho )\cdot \rho \cdot \delta ^{2}$ where $E_{%
\text{sym}}^{pot}(\rho )$ is given by Eq. (\ref{esympot}) from the MDI
interaction. Therefore, the momentum-independent SBKD potential that has $%
K_{0}=200$ MeV \cite{Ber84} and exactly the same $E_{\text{sym}}(\rho )$ as
the MDI interaction is
\begin{eqnarray}
U_{\text{SBKD}}(\rho ,\delta ,\tau )&=&-356~\rho /\rho
_{0}+303~(\rho/\rho _{0})^{7/6}  \notag \\
&+&4\tau E_{\text{sym}}^{\mathrm{pot}}(\rho )+(18.6-F(x))  \notag \\
&\times&(G(x)-1)(\rho /\rho _{0})^{G(x)}\delta ^{2}.  \label{Usbkd}
\end{eqnarray}

\subsection{In-medium nucleon-nucleon cross sections}

Another important quantity in the IBUU model is the in-medium NN
cross sections. While much attention has been given to finding
experimental observables that can constrain the symmetry energy,
little effort has been made so far to study the NN cross sections in
isospin asymmetric nuclear matter. Most of existing works on
in-medium NN cross sections have concentrated on their density
dependence in isospin symmetric nuclear matter, see, e.g.,
\cite{Neg81,Pan91,Li93,Sch97,Per02,Gia96,Koh98,LiQF00,Che01,Dan02b}.
One simple model for NN in-medium cross sections is the effective
mass scaling model \cite{Neg81,Pan91,Per02}. In this model, while
both the incoming current in the initial state and the level density
of the final state in an NN scattering depend on the effective
masses of colliding nucleons, the scattering matrix elements are
assumed to be the same in free-space and in the medium. As a result,
the ratio of the NN cross section in nuclear medium $\sigma
_{NN}^{\rm medium}$ to that in free-space value $\sigma _{NN}^{\rm
free}$ is simply given by
\begin{equation}
R_{\rm medium}\equiv \sigma _{NN}^{\rm medium}/\sigma _{NN}^{\rm
free}=(\mu _{NN}^{\ast }/\mu _{NN})^{2},  \label{xmedium}
\end{equation}%
where $\mu _{NN}$ and $\mu _{NN}^{\ast }$ are the reduced masses of
scattering nucleon pairs in free-space and in the medium,
respectively. Since the nucleon mass becomes smaller in nuclear
medium, the in-medium NN cross section is smaller than its value in
free space. The relation given in Eq. (\ref{xmedium}) was recently
found to be consistent with calculations based on the DBHF theory
\cite{Sam05b} for nucleon pairs with relative momenta less than
about 240 MeV/c and in nuclear matter with densities less than about
$2\rho _{0}$.  This finding thus lends a strong support to the
effective mass scaling model of in-medium NN cross sections in this
limited density and momentum ranges. We have thus extended this
model to determine the in-medium NN cross sections in the asymmetric
nuclear matter produced in nuclear reactions induced by radioactive
beams. For nucleon-nucleon scatterings at higher energies, inelastic
reaction channels become important. Although there were some studies
on in-medium effects in these channels \cite {Ber88a,Mao94,Gai05},
the experimental free-space cross sections are used in the IBUU
model as the model is mainly for heavy ion collisions at
intermediate energies where NN inelastic scatterings are less
important than elastic scatterings.

While the effective masses and the in-medium NN cross sections have
to be calculated dynamically in the evolving environment created
during heavy-ion reactions, it is instructive to examine the
in-medium NN cross sections in isospin asymmetric nuclear matter at
zero temperature. In this situation, the integrals in Eq.
(\ref{mdi}) can be analytically carried out. Specifically, it is
given by \cite{Das03,Pra88},
\begin{eqnarray}
&&\int d^{3}p^{\prime }\frac{f_{\tau }(\vec{r},\vec{p}^{\prime })}{1+(\vec{p}-%
\vec{p}^{\prime })^{2}/\Lambda ^{2}}\nonumber\\
&=&\frac{2}{h^{3}}\pi \Lambda ^{3}\left[ \frac{p_{f}^{2}(\tau
)+\Lambda ^{2}-p^{2}}{2p\Lambda }\ln\frac{(p+p_{f}(\tau
))^{2}+\Lambda ^{2}}{(p-p_{f}(\tau ))^{2}+\Lambda^{2}}\right.  \nonumber \\
&+&\left.\frac{2p_{f}(\tau )}{\Lambda}-2\tan^{-1}\frac{p+p_{f}(\tau
) }{\Lambda }-2 \tan^{-1}\frac{p-p_{f}(\tau )}{\Lambda }\right]
.\nonumber\\
\end{eqnarray}%
The reduction factor $R_{\rm medium}$ for in-medium NN cross
sections can thus also be obtained analytically, albeit lengthy.

\begin{figure}[tbh]
\includegraphics[height=0.35\textheight,angle=-90]{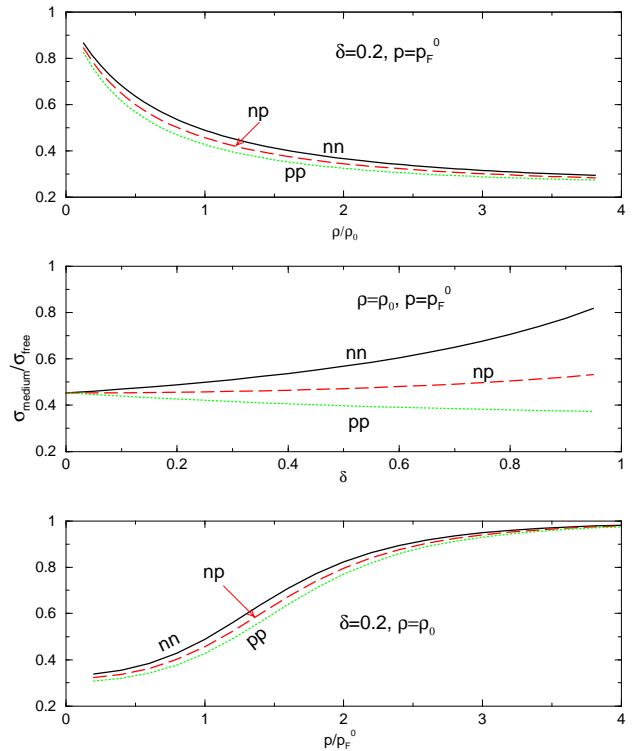}
\caption{(Color online) The reduction factor of the in-medium
nucleon-nucleon cross sections compared to their free-space values
as a function of density (top panel), isospin asymmetry (middle
panel) and momentum (bottom panel). Taken from Ref.
\protect\cite{LiBA05c}.} \label{medcrsc}
\end{figure}

As an illustration, we show in Fig.~\ref{medcrsc} the reduction
factor $R_{\rm medium}$ for a simplified case of two colliding
nucleons having the same momentum $p$. The $R_{\rm medium}$ factor
is examined as a function of density (top panel), isospin asymmetry
(middle panel) and the momentum (bottom panel). It is interesting to
note that not only in-medium NN cross sections are reduced compared
with their free-space values but also the nn and pp cross sections
become different while their free-space cross sections are the same.
Moreover, the difference between the nn and pp scattering cross
sections grows in more asymmetric matter. The larger in-medium cross
sections for nn than for pp are completely due to the positive
neutron-proton effective mass splitting with the effective
interaction used. This feature may serve as a probe of the
neutron-proton effective mass splitting in neutron-rich matter. We
also note that in-medium NN cross sections are also independent of
the parameter $x$ as they are solely determined by the momentum
dependence of the nuclear potential used in the model.

\section{Constraining the symmetry energy at sub-normal densities using
isospin diffusion data}

\label{diffusion}

The IBUU model outlined above have been used to study many
observables in heavy-ion reactions induced by both stable and
radioactive beams. In this section, we illustrate one of its
applications in extracting the density dependence of nuclear
symmetry energy from the isospin diffusion data in heavy-ion
collisions.

Isospin diffusion in heavy-ion collisions has been shown to depend
sensitively on the density dependence of nuclear symmetry energy
\cite{Far91,Shi03,LiBA04b}. Within a momentum-independent
transport model, in which the nuclear potential depends only on
local nuclear density, the isospin diffusion data from recent
experiments at the NSCL/MSU (National Superconducting Cyclotron
Laboratory at Michigan State University) was found to favor a
quadratic density dependence for the interaction part of nuclear
symmetry energy \cite{Tsa04}. This conclusion has stimulated much
interest because of its implications to nuclear many-body theories
and nuclear astrophysics. However, the nuclear potential acting on
a nucleon is known to depend also on its momentum. For nuclear
isoscalar potential, its momentum dependence is well-known and is
important in extracting information about the equation of state of
symmetric nuclear matter
\cite{Gal87,Wel88,Gal90,Pan93,Zha94,Gre99,Dan00,Per02,Dan02a}. The
momentum dependence of the isovector (symmetry) potential
\cite{Bom01,Hod94,Das03,LiBA04d} has also been shown to be
important for understanding a number of isospin related phenomena
in heavy-ion reactions \cite{LiBA04a,Riz04,Che04}. It is therefore
necessary to include momentum dependence in both isoscalar and
isovector potentials for studying the effect of nuclear symmetry
energy on isospin diffusion.

Isospin diffusion in heavy ion collisions can in principle be studied by
examining the average isospin asymmetry of the projectile-like residue in
the final state. Since reactions at intermediate energies are complicated by
preequilibrium particle emission and production of neutron-rich fragments at
mid-rapidity, differences of isospin diffusions in mixed and symmetric
systems are usually used to minimize these effects \cite{Tsa04}. To study
isospin diffusion in $^{124}$Sn + $^{112}$Sn reactions at $E=50$ \textrm{%
MeV/nucleon} and an impact parameter of $b=6$ fm, we thus also
consider the reaction systems $^{124}$Sn + $^{124}$S and
$^{112}$Sn + $^{112}$Sn and at same energy and impact parameter as
in Ref.~\cite{Tsa04}. The degree of isospin diffusion in the
reaction $^{124}$Sn + $^{112} $Sn is then measured by \cite{Ram00}
\begin{equation}
R_{i}=\frac{2X_{^{124}\text{Sn}+^{112}\text{Sn}}-X_{^{124}\text{Sn}+^{124}%
\text{Sn}}-X_{^{112}\text{Sn}+^{112}\text{Sn}}}{X_{^{124}\text{Sn}+^{124}%
\text{Sn}}-X_{^{112}\text{Sn}+^{112}\text{Sn}}}  \label{Ri}
\end{equation}%
where $X$ is the average isospin asymmetry $\left\langle \delta
\right\rangle $ of the $^{124}$Sn-like residue defined as the composition of
nucleons with local densities higher than $\rho _{0}/20$ and velocities
larger than $1/2$ the beam velocity in the c.m. frame. A density cut of $%
\rho _{0}/8$ is found to give almost same results. In ideal case, the value
of $R_{i}$ ranges between $0.05$ and $1$ from complete mixing to full
transparency.

\subsection{Effects of momentum-dependent interactions on isospin diffusion}

\begin{figure*}[tbh]
\includegraphics[scale=1.5]{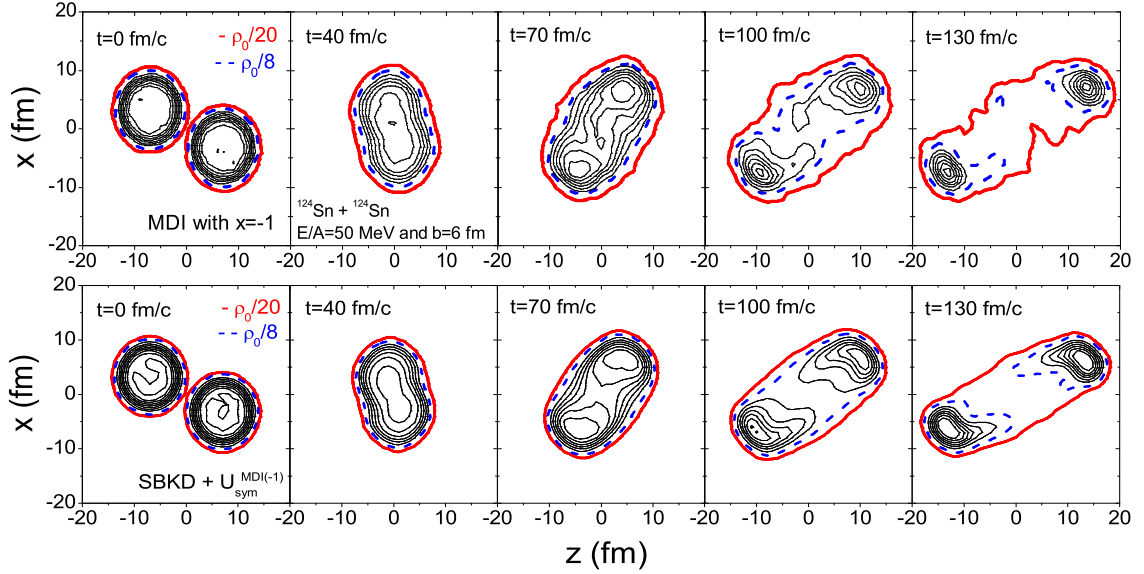}
\caption{{\protect\small (Color online) Density contour }$\protect\rho %
(x,0,z)${\protect\small \ in the reaction plane at different times
for the reaction }$^{124}${\protect\small Sn +
}$^{112}${\protect\small Sn at }$E/A=50${\protect\small \ MeV and
}$b=6${\protect\small \ fm by using momentum-dependent interaction
MDI with }$x=-1${\protect\small \ (upper panels) and
momentum-independent interaction SBKD with momentum-independent
symmetry potential }$U_{\text{sym}}^{\text{MDI}(-1)}(\protect\rho ,\protect%
\delta ,\protect\tau )${\protect\small \ (lower panels). Thick solid
lines represent }$\protect\rho _{{\protect\small 0}}${\protect\small
/20 while dashed lines represent }$\protect\rho _{{\protect\small 0}}$%
{\protect\small /8.}}
\label{CouXZ}
\end{figure*}

Effects of momentum-dependent interactions on dynamics of heavy-ion
collisions can be seen from the time evolution of the density
distribution \cite{Che04}. In Fig. \ref{CouXZ}, we show the density
contour $\rho (x,0,z)$ in the reaction plane at different times for
the reaction $^{124}$Sn + $^{112}$Sn at $E/A=50$ MeV and $b=6 $ fm
calculated with $x=-1$ using both the MDI and the soft Bertsch-Das
Gupta-Kruse (SBKD) interactions. It should be noted that the former
(MDI) interactions are momentum dependent for both isoscalar and
isovector nuclear potentials while the latter (SBKD) interactions do
not include any momentum dependence in either isoscalar or isovector
nuclear potentials though both interactions have similar
incompressibility $K_{0}$ and the same density dependence in the
symmetry energy. The experimental free space \textsl{N}-\textsl{N}
cross sections are used in these calculations. It is seen that both
interactions give similar time evolution of the collisions dynamics,
namely, projectile-like and target-like residues can be clearly
separated after about $100$ fm/c. Detailed examinations indicate,
however, that the reaction system expands more quickly and there are
also more emitted nucleons in the case of the momentum-dependent MDI
interaction than that of the momentum-independent SBKD interaction.

\begin{figure}[tbh]
\includegraphics[scale=0.85]{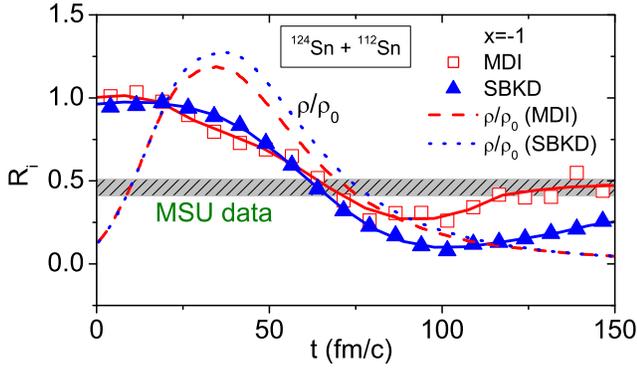}
\caption{(Color online) Degree of isospin diffusion as a function of
time with the MDI and SBKD interactions. Corresponding time
evolutions of central density are also shown. Taken from Ref.
\protect\cite{Che05a}.} \label{RiTime}
\end{figure}

\begin{figure}[tbh]
\includegraphics[scale=0.9]{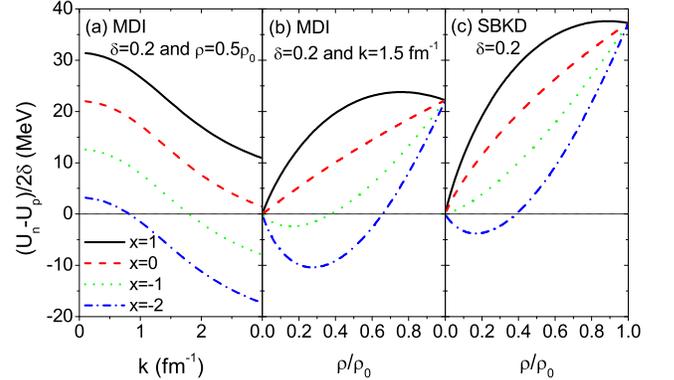}
\caption{(Color online) Symmetry potential as a function of nucleon
momentum (a) or density (b) with the MDI interaction and SBKD
interaction (c). Taken from Ref. \protect\cite{Che05a}.}
\label{Upot}
\end{figure}

In Fig.~\ref{RiTime}, we show the measured $R_i$ together with the
predictions from the IBUU04 for the time evolution of both $R_{i}$
and the average central densities calculated with $x=-1$ using
both the MDI and the soft SBKD interactions. It is seen that
isospin diffusion occurs mainly from about $30$ fm/c to $80$ fm/c
when average central density decreases from about $1.2\rho _{0}$
to $0.3\rho _{0}$. However, the value of $R_{i}$ still changes
slightly with time until after about $120$ fm/c when
projectile-like and target-like residues are well separated as
shown in Fig.~\ref{CouXZ}. This is partly due to the fact that the
isovector potential remains appreciable at low density as shown in
Fig.\ \ref{Upot}, where the symmetry potential
$(U_{n}-U_{p})/2\delta $ is shown as a function of nucleon
momentum (panel (a)) or density (panel (b)) for the MDI
interaction and as a function of density for the SBKD interaction
(panel (c)). Also, evaluating isospin diffusion $R_i$ based on
three reaction systems, which have different time evolutions for
the projectile residue as a result of different total energies and
numbers of nucleons, further contributes to the change of $R_i$ at
low density. For the two interactions consider here, the main
difference between the values for $R_i$ appears in the expansion
phase when densities in the participant region are well below
$\rho _{0}$. The experimental data from MSU are seen to be
reproduced nicely by the MDI interaction with $x=-1$, while the
SBKD interaction with $x=-1$ leads to a significantly lower value
for $R_{i}$ as the strength of the momentum-independent symmetry
potential is stronger (see Fig.\ \ref{Upot}), which has been shown
to enhance the isospin diffusion \cite{Far91,LiBA04b,Tsa04}.

\begin{figure}[tbh]
\includegraphics[scale=0.9]{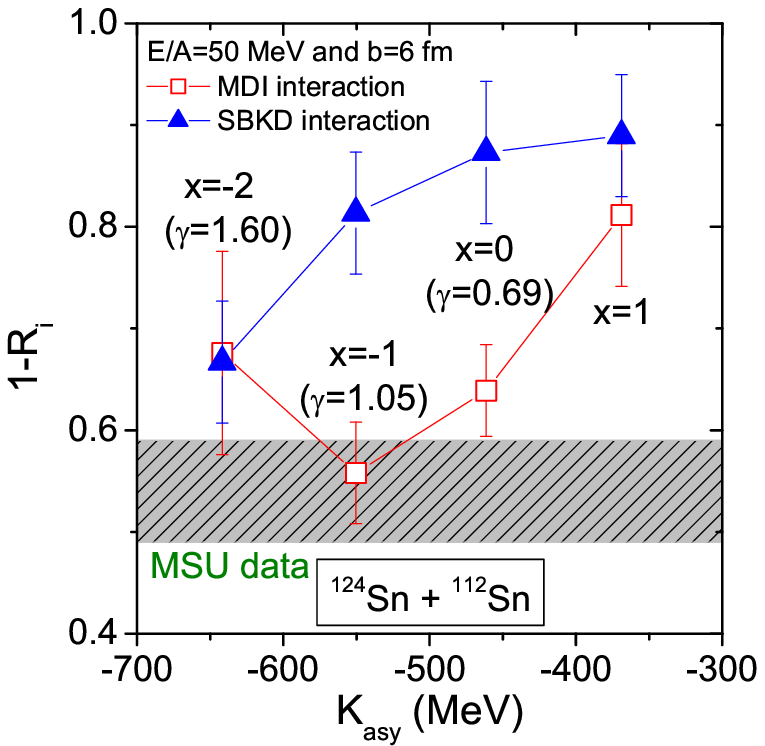}
\caption{(Color online) Degree of isospin diffusion as a function of $K_{%
\text{asy}}$ with the MDI and SBKD interactions. $\protect\gamma $ is the
parameter for fitting the corresponding symmetry energy with $E_{\text{sym}}(%
\protect\rho )=31.6(\protect\rho /\protect\rho _{0})^{\protect\gamma }$.
Taken from Ref. \protect\cite{Che05a}.}
\label{RiKasy}
\end{figure}

Effects of the symmetry energy on isospin diffusion were also
studied by varying the parameter $x$ \cite{Che05a}. Shown in
Fig.~\ref{RiKasy} is the final saturated value for $1-R_{i}$,
which measures the degree of isospin diffusion, as a function of
$K_{\text{asy}}$ for both MDI and SBKD interactions. It is
obtained by averaging the value of $1-R_{i}$ after $120$ fm/c with
error bars corresponding to its dispersion, whose magnitude is
similar to the error band shown in Ref. \cite{Tsa04} for the
theoretical results from the momentum-independent BUU model. For
the SBKD interaction without momentum dependence, the isospin
diffusion decreases monotonically (i.e., increasing value for
$R_i$) with increasing strength of $K_{\mathrm{asy}}$ as the
corresponding isovector potential is mostly positive and decreases
with increasing stiffness of $E_{\mathrm{sym}}(\rho)$ in the whole
range of considered $x$ parameter. The isospin diffusion is
reduced when the momentum-dependent interaction MDI is used as the
momentum dependence weakens the strength of symmetry potential
except for $x=-2$. As seen in Fig.~\ref{Upot}, the symmetry
potential in the MDI interaction has the smallest strength for
$x=-1$ as it is close to zero at $k\approx 1.5$ fm$^{-1}$ and
$\rho /\rho _{0}\approx 0.5$, and increases again with further
hardening of the symmetry energy, e.g., $x=-2$, when it becomes
largely negative at all momenta and densities. The MDI interaction
with $x=-1$ thus gives the smallest degree of isospin diffusion
among the interactions considered here and reproduces the MSU data
as already shown in Fig.~\ref{RiTime}.

The symmetry energy in the MDI interaction with $x=-1$ can be
parameterized as $E_{\rm sym}(\rho )\approx 31.6(\rho /\rho
_{0})^{1.05}$. It leads to a value of $K_{\mathrm{asy}}\approx -550$
MeV for the isospin dependent part of the isobaric incompressibility
of asymmetric nuclear matter, which should be compared to the
published constraint of $-566\pm 1350<K_{\mathrm{asy}}<139\pm 1617$
MeV extracted earlier from studying giant monopole resonances
\cite{Shl93}.

\subsection{Effects of in-medium NN cross sections on isospin diffusion}

The isospin degree of freedom plays an important role in heavy-ion
collisions through both the nuclear EOS and the nucleon-nucleon (NN)
scatterings \cite{LiBA98,LiBA01b}. In particular, the transport of
isospin asymmetry between two colliding nuclei is expected to depend
on not only the symmetry potential through the density dependence of
the symmetry energy $E_{\text{sym}}(\rho )$ but also the in-medium
NN cross sections. The two are related, respectively, to the
long-range and the short-range parts of the isospin-dependent
in-medium nuclear effective interactions. For instance, the drifting
contribution to the isospin transport in a nearly equilibrium system
is proportional to the product of the mean relaxation time $\tau
_{np}$ and the isospin asymmetric force $F_{np}$ \cite{Shi03}. While
$F_{np}$ is directly related to the gradient of the symmetry
potential, $\tau _{np}$ is inversely proportional to the
neutron-proton (np) scattering cross section $\sigma _{np}$
\cite{Shi03}. Furthermore, the collisional contribution to the
isospin transport in non-equilibrium system is generally expected to
be proportional to the np scattering cross section. In the above
study on isospin diffusion, the free-space NN cross sections were
used. In this section, we review the effects of in-medium NN cross
sections on the isospin diffusion in heavy ion collisions.

\begin{figure}[tbh]
\includegraphics[scale=0.85]{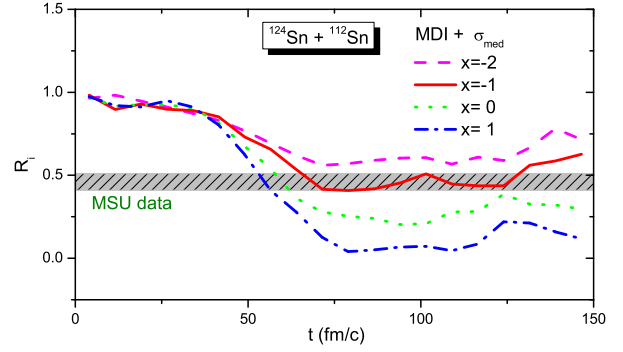}
\caption{(Color online) Time evolution of the isospin diffusion
$R_{i}$ using for MDI interactions with different $x$ parameters and
the in-medium nucleon-nucleon cross sections. Taken from Ref.
\protect\cite{LiBA05c}.} \label{RiTimeMed}
\end{figure}

Shown in Fig.~\ref{RiTimeMed} are the time evolution of $R_{i}$
re-calculated using the in-medium NN cross sections and the MDI
interaction with four $x$ parameters. It is seen that the net
isospin transport and the influence of the $x$ parameter show up
mainly in the expansion phase of the reaction after about $40$ fm/c
and become relatively stable after about $80$ fm/c. In the late
stage of the reaction, the values of $R_{i}$ from $x=-1$ and $0$
come close to the experimental data from MSU, which is shown by the
shaded band.

\begin{figure}[tbh]
\includegraphics[scale=0.90]{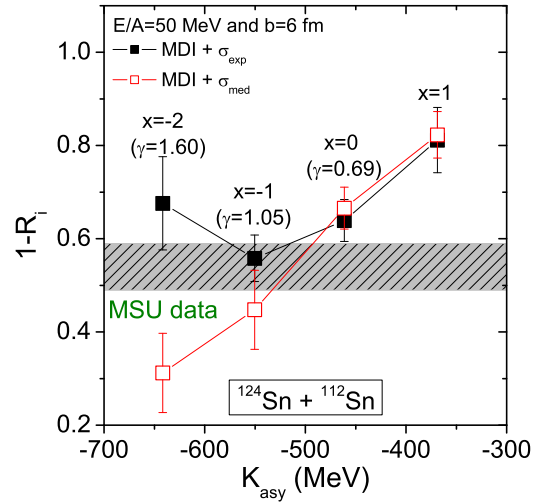}
\caption{(Color online) Degree of isospin diffusion as a function of $K_{%
\text{asy}}(\protect\rho _{0})$ with the free (filled squares) and in-medium
(open squares) nucleon-nucleon cross sections. Taken from Ref. \protect\cite%
{LiBA05c}.}
\label{RiKasyMed}
\end{figure}

For a more meaningful comparison with the experimental data, the
time average of $R_{i}$ between $t=120$ fm/c and $150$ fm/c has been
used as in Fig.~\ref{RiTime}. Shown in Fig.~\ref{RiKasyMed} is a
comparison of the averaged strength of isospin transport $1-R_{i}$
obtained with the free and in-medium NN cross sections,
respectively, as a function of the asymmetric part of the isobaric
incompressibility of nuclear matter at $\rho _{0}$, i.e.,
$K_{asy}(\rho _{0})$. The error bars were drawn to indicate
fluctuations and were obtained from the dispersion in the time
evolution of $R_{i}$ \cite{Che05a}. First, it is interesting to note
that with the in-medium NN cross sections the strength of isospin
transport $1-R_{i}$ decreases monotonically with decreasing value of
$x$. With the free-space NN cross sections, there appears, however,
to be a minimum at around $x=-1$. Moreover, this minimum is the
point closest to the experimental data. This allowed us to extract
the value of $K_{asy}(\rho _{0})=-550\pm 100$ MeV. With the
in-medium NN cross sections we can now further narrow down the
$K_{asy}(\rho _{0})$ to be about $-500\pm 50$ MeV. The latter is
consistent with that extracted from studying the isospin dependence
of giant resonances of $^{112} $Sn to $^{124}$Sn isotopes by
Fujiwara et al at Osaka \cite{Fuj05}. Also shown in the figure are
the $\gamma $ values used in fitting the symmetry energy with
$E_{\text{sym}}(\rho )=31.6(\rho /\rho _{0})^{\gamma }$. The results
with the in-medium NN cross sections constrain the $\gamma $
parameter to be between $0.69$ and $1.05$. The lower value is close
to what is extracted from studying giant resonances
\cite{Pie05,Col05}. The value of $\gamma =1.05$ extracted earlier
using the free-space NN cross sections sets an upper limit.

The difference in $1-R_{i}$ obtained with the free-space and the
in-medium NN cross sections is small for $x=1$ and $x=0$, but
becomes large for $x=-1$ and $x=-2$. The increasing effect of the
in-medium NN cross sections with decreasing $K_{asy}(\rho _{0})$ or
$x$ parameter can be understood from consideration of the
contributions from the symmetry potential and the np scatterings. As
we have mentioned above, both contributions to the isospin transport
depend on the np scattering cross section $\sigma _{np}$. While the
collisional contribution is proportional to the np scattering cross
section $\sigma _{np}$, the mean-field contribution is proportional
to the product of the isospin asymmetric force $F_{np}$ and the
inverse of $\sigma _{np}$. The overall effect of the in-medium NN
cross sections on isospin transport is a result of a complicated
combination of both the mean field and the NN scatterings. Generally
speaking, the symmetry potential effects on the isospin transport
will become weaker when the NN cross sections are larger while the
symmetry potential effects will show up more clearly if smaller NN
cross sections are used. This feature can be seen from
Fig.~\ref{Upot}. For $x=1$ and $x=0$, the symmetry potential and its
gradient with respect to density, as shown in Fig.~\ref{Upot}, are
large at low densities where the majority of net isospin transport
occurs. The $F_{np}$ factor makes the contribution due to the mean
field to dominate over that due to the collisions. Therefore, the
reduced in-medium $\sigma _{np}$ leads to about the same or a
slightly higher isospin transport. As the $x$ parameter decreases to
$x=-1$ and $x=-2$, however, the symmetry potential decreases and its
density gradient can be even negative at low densities. In these
cases, either the collisional contribution dominates or the
mean-field contribution becomes negative. Therefore, the reduced
in-medium np scattering cross section $\sigma _{np}$ leads to a
weaker isospin transport compared with the case with the free-space
NN cross sections.

\begin{figure}[tbp]
\includegraphics[scale=0.95]{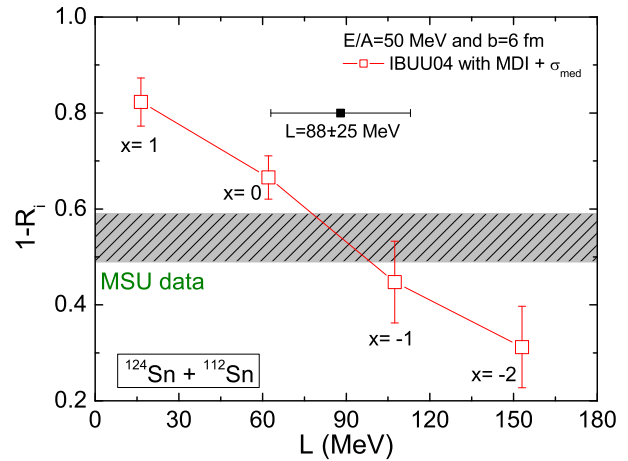}
\caption{(Color online) Degree of the isospin diffusion $1-R_{i}$ as a
function of $L$ using the MDI interaction with $x=-2$, $-1$, $0$, and $1$.
The shaded band indicates the data from NSCL/MSU \protect\cite{Tsa04}. The
solid square with error bar represents $L=88\pm 25$ MeV. Taken from Ref.
\protect\cite{Che05b}.}
\label{RiL}
\end{figure}

Based on an isospin- and momentum-dependent IBUU04 transport model with
free-space experimental NN cross sections, comparing the theoretical results
with the experimental data has allowed us to extract a nuclear symmetry
energy of $E_{\text{sym}}(\rho )\approx 31.6(\rho /\rho _{0})^{1.05}$.
Including also medium-dependent NN cross sections, which are important for
isospin-dependent observables \cite{LiBA05c,LiBA05d}, the isospin diffusion
data leads to an even softer nuclear symmetry energy of $E_{\text{sym}}(\rho
)\approx 31.6(\rho /\rho _{0})^{\gamma }$ with $\gamma \approx 0.69 $ \cite%
{LiBA05c}.

Since the slope parameter $L$ of the nuclear symmetry energy gives an
important constraint on the density dependence of the nuclear symmetry
energy, in particular, it has been shown that the slope parameter $L$ is
related to the neutron skin thickness of heavy nuclei, it is interesting to
see how the isospin diffusion data constrain the value of $L$. In Fig.~\ref%
{RiL}, we show the results from the IBUU04 transport model with in-medium NN
cross sections, that are consistent with the mean-field potential obtained
with the MDI interactions used in the model, for the degree of the isospin
diffusion $1-R_{i}$ as a function of $L$. The shaded band in Fig. \ref{RiL}
indicates the data from NSCL/MSU \cite{Tsa04}. It is seen that the strength
of isospin diffusion $1-R_{i}$ decreases monotonically with decreasing value
of $x$ or increasing value of $L$. This is expected as the parameter $L$
reflects the difference in the pressures on neutrons and protons. From
comparison of the theoretical results with the data, we can clearly exclude
the MDI interaction with $x=1$ and $x=-2$ as they give either too large or
too small a value for $1-R_{i}$ compared to that of data. The range of $x$
or $L$ values that give values of $1-R_{i}$ falling within the band of
experimental values could in principle be determined in our model by
detailed calculations. Instead, we determine this schematically by using the
results from the four $x$ values. For the centroid value of $L$, it is
obtained from the interception of the line connecting the theoretical
results at $x=-1$ and $0$ with the central value of $1-R_{i}$ data in Fig. %
\ref{RiL}, i.e., $L=88$ MeV. The upper limit of $L=113$ MeV is estimated
from the interception of the line connecting the upper error bars of the
theoretical results at $x=-1$ and $-2$ with the lower limit of the data band
of $1-R_{i}$. Similarly, the lower limit of $L=65$ MeV is estimated from the
interception of the line connecting the lower error bars of the theoretical
results at $x=0$ and $-1$ with the upper limit of the data band of $1-R_{i}$%
. This leads to an extracted value of $L=88\pm 25$ MeV as shown by the solid
square with error bar in Fig. \ref{RiL}.

\section{Constraining the Skyrme effective interactions and the neutron skin
thickness of nuclei using isospin diffusion data from heavy ion collisions}

\label{skyrmenskin}

Information on the density dependence of the nuclear symmetry
energy can in principle also be obtained from the thickness of the
neutron skin in heavy nuclei as the latter is strongly correlated
with the slope parameter $L$ of the
nuclear symmetry energy at saturation density \cite%
{Bro00,Hor01a,Typ01,Fur02,Kar02,Die03}. Because of the large uncertainties
in measured neutron skin thickness of heavy nuclei, this has not been
possible. Instead, studies have been carried out to use the extracted
nuclear symmetry energy from the isospin diffusion data to constrain the
neutron skin thickness of heavy nuclei \cite{Ste05b,LiBA05c}. Using the
Hartree-Fock approximation with parameters fitted to the phenomenological
EOS that was used in the IBUU04 transport model to describe the isospin
diffusion data from the NSCL/MSU, it was found that a neutron skin thickness
of less than $0.15$ fm \cite{Ste05b,LiBA05c} for $^{208}$Pb was incompatible
with the isospin diffusion data.

In the following, we study more systematically the correlation between the
density dependence of the nuclear symmetry energy and the thickness of the
neutron skin in a number of nuclei within the framework of the Skyrme
Hartree-Fock model. Using the extracted values of $L$ from the isospin
diffusion data in heavy-ion collisions, we obtain stringent constraints on
the neutron skin thickness of the nuclei $^{208}$Pb, $^{132}$Sn, and $^{124}$%
Sn. The extracted value of $L$ also limits the allowed parameter sets for
the Skyrme interaction.

\subsection{Constraining the Skyrme effective interactions using isospin
diffusion data}

In the standard Skyrme Hartree-Fock model, the interaction is taken to have
a zero-range, density- and momentum-dependent form \cite%
{Bra85,Fri86,Bro98,Che99b,Sto03}, i.e.,
\begin{eqnarray}
V_{12}(\mathbf{R},\mathbf{r}) &=&t_{0}(1+x_{0}P_{\sigma })\delta (\mathbf{r})
\notag \\
&+&\frac{1}{6}t_{3}(1+x_{3}P_{\sigma })\rho ^{\sigma }(\mathbf{R})\delta (%
\mathbf{r})  \notag \\
&+&\frac{1}{2}t_{1}(1+x_{1}P_{\sigma })(K^{^{\prime }2}\delta (\mathbf{r}%
)+\delta (\mathbf{r})K^{2})  \notag \\
&+&t_{2}(1+x_{2}P_{\sigma })\mathbf{K}^{^{\prime }}\cdot \delta (\mathbf{r})%
\mathbf{K}  \notag \\
&\mathbf{+}&iW_{0}\mathbf{K}^{^{\prime }}\cdot \delta (\mathbf{r})[(\mathbf{%
\sigma }_{1}+\mathbf{\sigma }_{2})\times \mathbf{K]},  \label{Sky}
\end{eqnarray}%
with $\mathbf{r}=\mathbf{r}_{1}-\mathbf{r}_{2}$ and $\mathbf{R}=(\mathbf{r}%
_{1}+\mathbf{r}_{2})/2$. In the above, the relative momentum operators $%
\mathbf{K}=(\mathbf{\nabla }_{1}-\mathbf{\nabla }_{2})/2i$ and $\mathbf{K}%
^{\prime }=-(\mathbf{\nabla }_{1}-\mathbf{\nabla }_{2})/2i$ act on the wave
function on the right and left, respectively. The quantities $P_{\sigma }$
and $\sigma _{i}$ denote, respectively, the spin exchange operator and Pauli
spin matrices. The $\sigma$, $t_{0}-t_{3}$, $x_{0}-x_{3}$, and $W_{0}$ are
Skyrme interaction parameters that are chosen to fit the binding energies
and charge radii of a large number of nuclei in the periodic table. For
infinite nuclear matter, the symmetry energy from the Skyrme interaction can
be expressed as \cite{Che99b,Sto03}
\begin{eqnarray}
E_{\text{sym}}(\rho ) &=&\frac{1}{3}\frac{\hbar ^{2}}{2m}\left( \frac{3\pi
^{2}}{2}\right) ^{2/3}\rho ^{2/3}  \notag \\
&-&\frac{1}{8}t_{0}(2x_{0}+1)\rho -\frac{1}{48}t_{3}(2x_{3}+1)\rho ^{\sigma
+1}  \notag \\
&+&\frac{1}{24}\left( \frac{3\pi ^{2}}{2}\right) ^{2/3}\left[
-3t_{1}x_{1}\right.  \notag \\
&+&\left. \left( 4+5x_{2}\right) t_{2}\right] \rho ^{5/3}.  \label{EsymSky}
\end{eqnarray}%
The coefficient of the $\delta ^{4}$ term in Eq. (\ref{EsymPara}) can also
be obtained analytically and has been shown to be very small over a large
range of nuclear density ($\leq $ 0.80 fm$^{-3}$) and isospin asymmetry. The
parabolic law of Eq. (\ref{EsymPara}) without the $\delta ^{4}$ and
higher-order terms in $\delta $ is thus justified \cite{Che99b}.

\begin{figure}[tbp]
\includegraphics[scale=0.9]{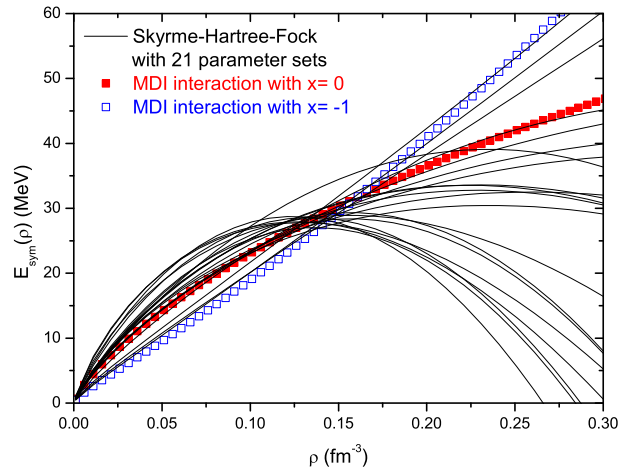}
\caption{(Color online) Density dependence of the nuclear symmetry energy $%
E_{\text{sym}}(\protect\rho )$ for 21 sets of Skyrme interaction parameters.
The results from the MDI interaction with $x=-1$ (open squares) and $0$
(solid squares) are also shown. Taken from Ref. \protect\cite{Che05b}.}
\label{SymDen}
\end{figure}

Fig. \ref{SymDen} displays the density dependence of $E_{\text{sym}}(\rho )$
for $21$ sets of Skyrme interaction parameters, i.e., $SKM$, $SKM^{\ast }$, $%
RATP$, $SI$, $SII$, $SIII$, $SIV$, $SV$, $SVI$, $E$, $E_{\sigma }$, $%
G_{\sigma }$, $R_{\sigma }$, $Z$, $Z_{\sigma }$, $Z_{\sigma }^{\ast }$, $T$,
$T3$, $SkX$, $SkXce$, and $SkXm$. The values of the parameters in these
Skyrme interactions can be found in Refs. \cite{Bra85,Fri86,Bro98}. For
comparison, we also show in Fig. \ref{SymDen} results from the
phenomenological MDI interactions with $x=-1$ (open squares) and $0$ (solid
squares). As we have discussed above, from comparing the isospin diffusion
data from NSCL/MSU using the IBUU04 with in-medium NN cross sections, these
interactions are recently shown to give, respectively, the upper and lower
bounds for the stiffness of the symmetry energy \cite{LiBA05c}. It is seen
from Fig. \ref{SymDen} that the density dependence of the symmetry energy
varies drastically among different interactions. Although the values of $E_{%
\text{sym}}(\rho _{0})$ are all in the range of $26$-$35$ MeV, the values of
$L$ and $K_{\text{sym}}$ are in the range of $-50$-$100$ MeV and $-700$-$50$
MeV, respectively.

The extracted value of $L=88\pm 25$ MeV gives a rather stringent constraint
on the density dependence of the nuclear symmetry energy and thus puts
strong constraints on the nuclear effective interactions as well. For the
Skyrme effective interactions shown in Fig. \ref{SymDen}, for instance, all
of those lie beyond $x=0$ and $x=-1$ in the sub-saturation region are not
consistent with the extracted value of $L$. Actually, we note that only $4$
sets of Skyrme interactions, i.e., $\mathrm{SIV}$, $\mathrm{SV}$, $\mathrm{G}%
_{\sigma }$, and $\mathrm{R}_{\sigma }$, in the $21$ sets of Skyrme
interactions considered here have nuclear symmetry energies that are
consistent with the extracted $L$ value.

\subsection{Constraining the neutron skin thickness of nuclei using isospin
diffusion data}

\begin{figure}[tbp]
\includegraphics[scale=0.9]{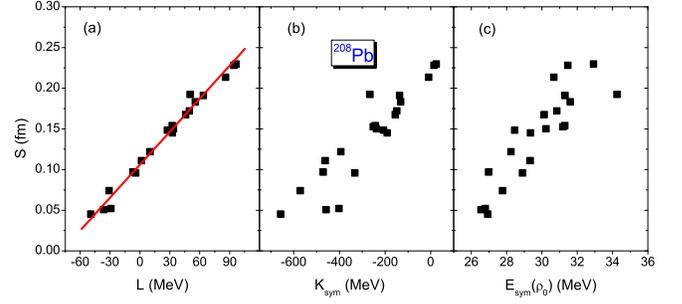}
\caption{(Color online) {Neutron skin thickness $S$ of $^{208}$Pb as a
function of (a) $L$, (b) $K_{\text{sym}}$, and (c) $E_{\text{sym}}(\protect%
\rho _{0})$ for 21 sets of Skyrme interaction parameters. }The line in panel
(a) represents a linear fit. Taken from Ref. \protect\cite{Che05b}.}
\label{SPb208}
\end{figure}

The neutron skin thickness $S$ of a nucleus is defined as the difference
between the root-mean-square radii $\sqrt{\left\langle r_{n}\right\rangle }$
of neutrons and $\sqrt{\left\langle r_{p}\right\rangle }$ of protons, i.e.,
\begin{equation}
S=\sqrt{\left\langle r_{n}^{2}\right\rangle }-\sqrt{\left\langle
r_{p}^{2}\right\rangle }.  \label{S}
\end{equation}%
It has been known that $S$ is sensitive to the density dependence of the
nuclear symmetry energy, particularly the slope parameter $L$ at the normal
nuclear matter density \cite{Bro00,Hor01a,Typ01,Fur02,Kar02,Die03}. Using
above $21$ sets of Skyrme interaction parameters, we have evaluated the
neutron skin thickness of several nuclei. In Figs. \ref{SPb208}(a), (b) and
(c), we show, respectively, the correlations between the neutron skin
thickness of $^{208}$Pb with $L$, $K_{\text{sym}}$, and $E_{\text{sym}}(\rho
_{0})$. It is seen from Fig. \ref{SPb208}(a) that there exists an
approximate linear correlation between $S$ and $L$. The correlations of $S$
with $K_{\text{sym}}$ and $E_{\text{sym}}(\rho _{0})$ are less strong and
even exhibit some irregular behavior. The solid line in Fig. \ref{SPb208}(a)
is a linear fit to the correlation between $S$ and $L$ and is given by the
following expression:
\begin{eqnarray}
S(^{\text{208}}\text{Pb)} &=&(0.1066\pm 0.0019)  \notag \\
&&+(0.00133\pm 3.76\times 10^{-5})\times L,  \label{SLPb208a}
\end{eqnarray}%
or
\begin{eqnarray}
L &=&(-78.5\pm 3.2)  \notag \\
&&+(740.4\pm 20.9)\times S(^{\text{208}}\text{Pb)},  \label{SLPb208b}
\end{eqnarray}%
where the units of $L$ and $S$ are \textrm{MeV} and \textrm{fm},
respectively. Therefore, if the value for either $S(^{\text{208}}$Pb) or $L$
is known, the value for the other can be determined.

\begin{figure}[tbp]
\includegraphics[scale=0.9]{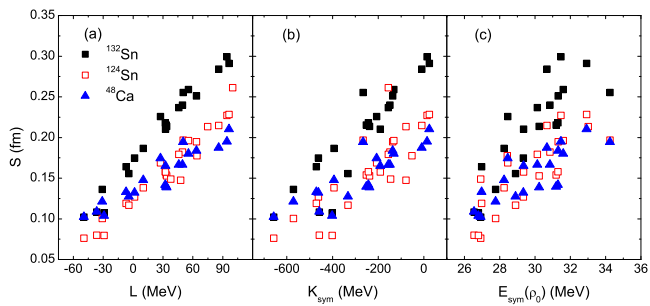}
\caption{(Color online) {Same as Fig. 2 but for nuclei $^{132}$Sn (Solid
squares), $^{124}$Sn (Open squares) and $^{48}$Ca (Triangles).} Taken from
Ref. \protect\cite{Che05b}.}
\label{SSnCa}
\end{figure}

It is of interest to see if there are also correlations between the neutron
skin thickness of other neutron-rich nuclei and the nuclear symmetry energy.
Fig. \ref{SSnCa} shows the same correlations as in Fig. \ref{SPb208} but for
the neutron-rich nuclei $^{132}$Sn, $^{124}$Sn, and $^{48}$Ca. For the heavy
$^{132}$Sn and $^{124}$Sn, we obtain a similar conclusion as for $^{208}$Pb,
namely, $S$ exhibits an approximate linear correlation with $L$ but weaker
correlations with $K_{\text{sym}}$ and $E_{\text{sym}}(\rho _{0})$. For the
lighter $^{48}$Ca, on the other hand, all the correlations become weaker
than those of heavier nuclei. Therefore, the neutron skin thickness of heavy
nuclei is better correlated with the density dependence of the nuclear
symmetry energy. As in Eqs. (\ref{SLPb208a}) and (\ref{SLPb208b}), a linear
fit to the correlation between $S$ and $L$ can also be obtained for $^{132}$%
Sn and $^{124}$Sn, and the corresponding expressions are
\begin{eqnarray}
S(^{\text{132}}\text{Sn)} &=&(0.1694\pm 0.0025)  \notag \\
&&+(0.0014\pm 5.12\times 10^{-5})\times L,  \label{SLSn132a}
\end{eqnarray}%
\begin{eqnarray}
L &=&(-117.1\pm 5.4)  \notag \\
&&+(695.1\pm 25.3)\times S(^{\text{132}}\text{Sn)},  \label{SLSn132b}
\end{eqnarray}%
and
\begin{eqnarray}
S(^{\text{124}}\text{Sn)} &=&(0.1255\pm 0.0020)  \notag \\
&&+(0.0011\pm 4.05\times 10^{-5})\times L. \label{SLSn124a}
\end{eqnarray}%
\begin{eqnarray}
L &=&(-110.1\pm 5.2)  \notag \\
&&+(882.6\pm 32.3)\times S(^{\text{124}}\text{Sn)},  \label{SLSn124b}
\end{eqnarray}%

Similar linear relations between $S$ and $L$ are also expected for other
heavy nuclei. This is not surprising as detailed discussions in Refs. \cite%
{Bro00,Hor01a,Typ01,Fur02,Kar02,Die03} have shown that the thickness of the
neutron skin in heavy nuclei is determined by the pressure difference
between neutrons and protons, which is proportional to the parameter $L$.

\begin{table}[tbp]
\caption{{\protect\small {Linear correlation coefficients $C_{l}$ of $S$
with $L$, $K_{\text{sym}}$ and $E_{\text{sym}}(\protect\rho _{0})$ for $%
^{208}$Pb,\ $^{132}$Sn, $^{124}$Sn, and $^{48}$Ca from 21 sets of Skyrme
interaction parameters.} Taken from Ref. \protect\cite{Che05b}.}}
\label{Corr}%
\begin{tabular}{ccccc}
\hline\hline
$C_{l}$ $(\%)$ & \quad $^{208}$Pb\quad & $\quad ^{132}$Sn \quad & $^{124}$Sn
& $^{48}$Ca \\ \hline
$S$-$L$ & $99.25$ & $98.76$ & $98.75$ & $93.66$ \\
$S$-$K_{\text{sym}}$ & $92.26$ & $92.06$ & $92.22$ & $86.99$ \\
$S$-$E_{\text{sym}}$ & $87.89$ & $85.74$ & $85.77$ & $81.01$ \\ \hline\hline
\end{tabular}%
\end{table}

To give a quantitative estimate of above discussed correlations, we define
the following linear correlation coefficient $C_{l}$:
\begin{equation}
C_{l}=\sqrt{1-q/t},
\end{equation}%
where%
\begin{eqnarray}
q &=&\underset{i=1}{\overset{n}{\sum }}[y_{i}-(A+Bx_{i})]^{2}, \\
t &=&\underset{i=1}{\overset{n}{\sum }}(y_{i}-\overline{y}),~~~\overline{y}=%
\underset{i=1}{\overset{n}{\sum }}y_{i}/n.
\end{eqnarray}%
In the above, $A$ and $B$ are the linear regression coefficients, $(x_{i}$, $%
y_{i})$ are the sample points, and $n$ is the number of sample points. The
linear correlation coefficient $C_{l}$ measures the degree of linear
correlation, and $C_{l}=1$ corresponds to an ideal linear correlation. Table %
\ref{Corr} gives the linear correlation coefficient $C_{l}$ for the
correlation of $S$ with $L$, $K_{\text{sym}}$ and $E_{\text{sym}}(\rho _{0})$
for $^{208}$Pb, $^{132}$Sn, $^{124}$Sn, and $^{48}$Ca shown in Figs. \ref%
{SPb208} and \ref{SSnCa} for different Skyrme interactions. It is seen that
these correlations become weaker with decreasing nucleus mass, and a strong
linear correlation only exists between the $S$ and $L$ for the heavier
nuclei $^{208}$Pb, $^{132}$Sn, and $^{124}$Sn. Therefore, the neutron skin
thickness of these nuclei can be extracted once the slope parameter $L$ of
the nuclear symmetry energy at saturation density is known.

The extracted $L$ value from isospin diffusion data allows us to
determine from Eqs. (\ref{SLPb208a}), (\ref{SLSn132a}), and
(\ref{SLSn124a}), respectively, a neutron skin
thickness of $0.22\pm 0.04$ fm for $^{208}$Pb, $0.29\pm 0.04$ fm for $^{132}$%
Sn, and $0.22\pm 0.04$ fm for $^{124}$Sn. Experimentally, great efforts were
devoted to measure the thickness of the neutron skin in heavy nuclei \cite%
{Sta94,Cla03}, and a recent review can be found in Ref. \cite{Kra04}. The
data for the neutron skin thickness of $^{208}$Pb indicate a large
uncertainty, i.e., $0.1$-$0.28$ fm. Our results for the neutron skin
thickness of $^{208}$Pb are thus consistent with present data but give a
much stronger constraint. A large uncertainty is also found experimentally
in the neutron skin thickness of $^{124}$Sn, i.e., its value varies from $%
0.1 $ fm to $0.3$ fm depending on the experimental method. The proposed
experiment of parity-violating electron scattering from $^{208}$Pb at the
Jefferson Laboratory is expected to give another independent and more
accurate measurement of its neutron skin thickness (within $0.05$ fm), thus
providing improved constraints on the density dependence of the nuclear
symmetry energy \cite{Hor01b,Jef00}.

Recently, an accurately calibrated relativistic parametrization based on the
relativistic mean-field theory has been introduced to study the neutron skin
thickness of finite nuclei \cite{Tod05}. This parametrization can describe
simultaneously the ground state properties of finite nuclei and their
monopole and dipole resonances. Using this parametrization, the authors
predicted a neutron skin thickness of $0.21$ fm in $^{208}$Pb, $0.27$ fm in $%
^{132}$Sn, and $0.19$ fm in $^{124}$Sn \cite{Tod05,Pie05}. These predictions
are in surprisingly good agreement with our results constrained by the
isospin diffusion data in heavy-ion collisions.

In addition, the neutron skin thickness of the nucleus $^{90}$Zr has
recently been determined to be $0.07\pm 0.04$ fm from the
model-independent spin-dipole sum rule value measured from the
charge-exchange spin-dipole excitations \cite{Yak06}. This value is
reproduced by the symmetry energy with $L=88\pm 25$ MeV extracted
from the isospin diffusion data in heavy-ion collisions, which
predicts a neutron skin thickness of $0.088\pm 0.04$ fm for
$^{90}$Zr.

\section{Probing the high density behavior of the nuclear symmetry energy in
heavy-ion reactions induced by high energy radioactive beams}

\label{highdensity}

Although significant progress has been made in the determination of
the density dependence of the nuclear symmetry energy at sub-normal
densities, the high density behavior of the the nuclear symmetry
energy is still poorly known. Fortunately, heavy-ion reactions,
especially those induced by high energy radioactive beams to be
available at high energy radioactive beam facilities, provide a
unique opportunity to pin down the high density behavior of the
symmetry energy. In this section, we illustrate via transport model
simulations several experimental observables which are sensitive to
the high density behavior of the symmetry energy.

\subsection{Isospin asymmetry of dense matter formed in high energy
heavy-ion reactions}

\begin{figure}[tbh]
\includegraphics[scale=0.5]{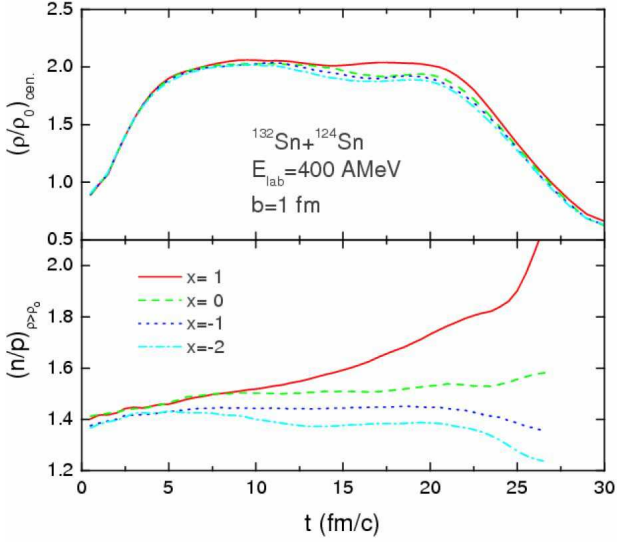}
\caption{(Color online) Central baryon density (upper window) and
isospin
asymmetry (lower window) of high density region for the reaction of $%
^{132}Sn+^{124}Sn$ at a beam energy of 400 MeV/nucleon and an impact
parameter of 1 fm. Taken from Ref. \protect\cite{LiBA05a}.}
\label{CentDen}
\end{figure}

To see the maximum baryon density and isospin asymmetry that can be
achieved in central heavy ion collisions induced by high energy
radioactive beams in future radioactive beam facilities, we show in
Fig.~\ref{CentDen} the central baryon density (upper window) and the
average $(n/p)_{\rho \geq \rho _{0}}$ ratio (lower window) of all
regions with baryon densities higher than $\rho _{0}$ in the
reaction of $^{132}$Sn$+^{124}$Sn at a beam energy of 400
MeV/nucleon and an impact parameter of 1 fm. It is seen that the
maximum baryon density is about 2 times normal nuclear matter
density. Moreover, the compression is rather insensitive to the
symmetry energy because the latter is relatively small compared to
the EOS of symmetric matter around this density. The high density
phase lasts for about 15 fm/c from 5 to 20 fm/c for this reaction.
The isospin asymmetry of the high density region is, however,
sensitive to the symmetry energy. The soft (e.g., $x=1$) symmetry
energy leads to a significantly higher value of $(n/p)_{\rho \geq
\rho _{0}}$ than the stiff one (e.g., $x=-2$). This is consistent
with the well-known isospin fractionation phenomenon that it is
energetically more favorable to have a higher isospin asymmetry
$\delta $ in the high density region for a softer symmetry energy
functional $E_{\rm sym}(\rho )$ as a result of the $E_{\rm sym}(\rho
)\delta ^{2}$ term in the EOS of asymmetric nuclear matter. Since
the symmetry energy changes from being soft to stiff when the
parameter $x$ varies from 1 to $-2$, the value of $(n/p)_{\rho \geq
\rho _{0}}$ becomes lower in the supranormal density region as the
parameter $x$ changes from 1 to $-2$. Because of neutron-skins of
the colliding nuclei, especially that of the projectile $^{132}{\rm
Sn}$, the initial value of the quantity $(n/p)_{\rho \geq \rho
_{0}}$, which is about 1.4, is less than the average n/p ratio of
1.56 of the reaction system. Also, in neutron-rich nuclei, the n/p
ratio on the low-density surface is much higher than that in their
interior. The dense matter region in heavy ion collisions can thus
become either neutron-richer or neutron-poorer with respect to the
initial state depending on the symmetry energy functional $E_{\rm
sym}(\rho )$ used.

\subsection{Isospin fractionation and n-p differential flow}

\begin{figure}[tbh]
\includegraphics[scale=0.5]{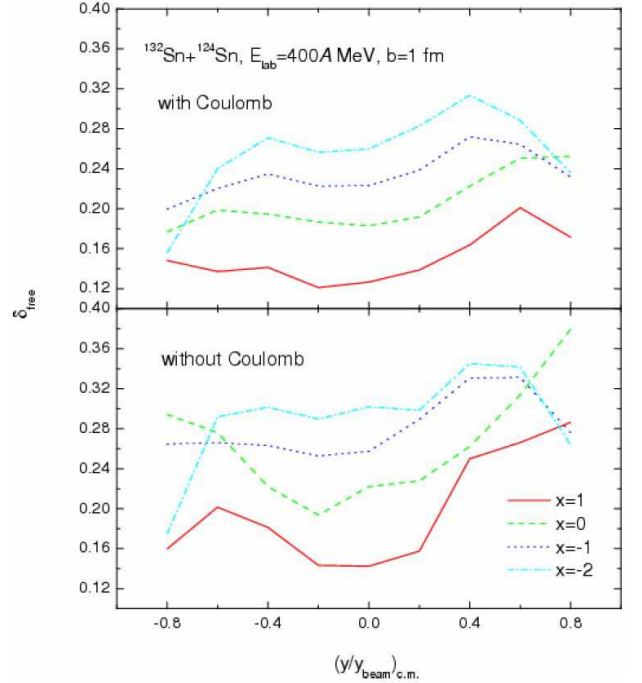}
\caption{(Color online) Isospin asymmetry of free nucleons with and
without the Coulomb force for different symmetry energies. Taken
from Ref. \protect\cite{LiBA05b}.} \label{deltay}
\end{figure}

The degree of isospin equilibration or translucency in heavy ion
collisions can be measured by the rapidity distribution of nucleon
isospin asymmetry $\delta _{\rm free}\equiv
(N_{n}-N_{p})/(N_{n}+N_{p})$, where $N_{n}$ and $N_{p}$ are
multiplicities of free neutrons and protons, respectively
\cite{LiBA04a}. Although it might be difficult to measure directly
$\delta _{\rm free}$ because it requires the detection of neutrons,
similar information can be extracted from ratios of light clusters,
such as, t/$^{3}$He, as demonstrated recently within a coalescence
model \cite{Che03b,Che04}. Shown in Fig.~\ref{deltay} are rapidity
distributions of $\delta _{\rm free}$ with (upper window) and
without (lower window) the Coulomb potential. It is interesting to
see that the $\delta _{\rm free}$ at midrapidity is particularly
sensitive to the symmetry energy. As the parameter $x$ increases
from $-2$ to $1$ the $\delta _{\rm free}$ at midrapidity decreases
by about a factor of 2. Moreover, the forward-backward asymmetric
rapidity distributions of $\delta _{\rm free}$ with all four $x$
parameters indicates the apparent nuclear translucency during the
reaction \cite{LiBA05b}.

\begin{figure}[tbh]
\includegraphics[scale=0.5]{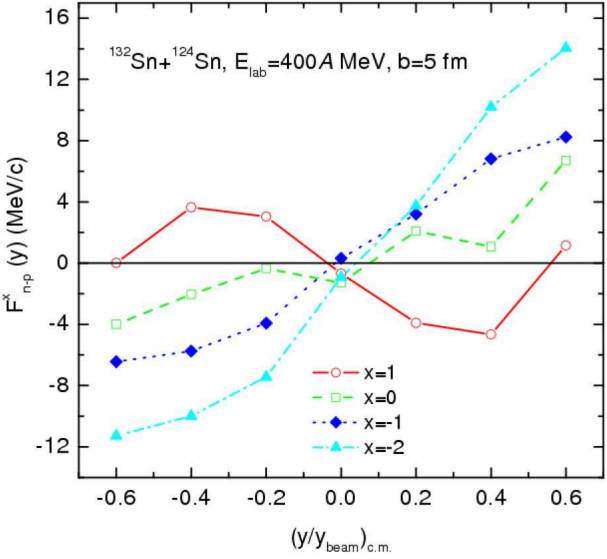}
\caption{(Color online) Neutron-proton differential flow for
different symmetry energies. Taken from Ref.
\protect\cite{LiBA05e}.} \label{npdiff}
\end{figure}

Another observable that is sensitive to the high density behavior of
the symmetry energy is the neutron-proton differential flow first
introduced in Ref.~\cite{LiBA00}
\begin{equation}
F_{n-p}^{x}(y)\equiv \sum_{i=1}^{N(y)}(p_{i}^{x}w_{i})/N(y),
\label{npdiffflow}
\end{equation}%
where $w_{i}=1(-1)$ for neutrons (protons) and $N(y)$ is the total
number of free nucleons at rapidity $y$. Since the differential flow
depends on the symmetry potential through the latter's effects on
the isospin fractionation and the collective flow, it has the
advantage of maximizing the effects of the symmetry potential while
minimizing those of the isoscalar potential. Shown in
Fig.~\ref{npdiff} is the n-p differential flow for the reaction of
$^{132} $Sn$+^{124}$Sn at a beam energy of 400 MeV/nucleon and an
impact parameter of 5 fm. Effects of the symmetry energy are clearly
revealed by changing the parameter $x$.

\subsection{Pion yields and $\protect\pi ^{-}/\protect\pi ^{+}$ ratio}

\begin{figure}[tbh]
\includegraphics[scale=0.5]{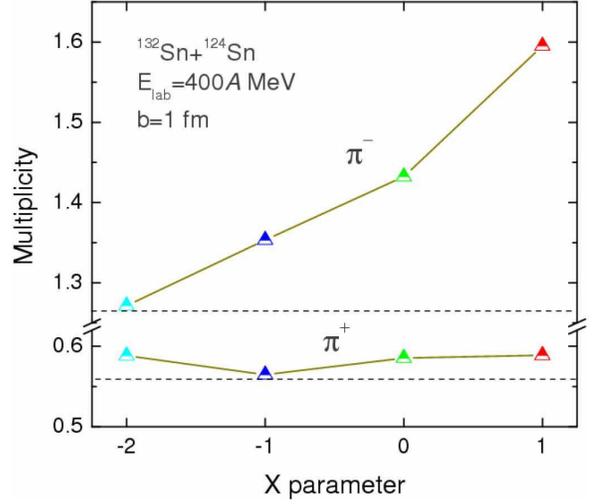}
\caption{(Color online) The $\protect\pi ^{-}$ and $\protect\pi
^{+}$ yields as functions of the $x$ parameter. Taken from Ref.
\protect\cite{LiBA05a}.} \label{pionyield}
\end{figure}

\begin{figure}[tbh]
\includegraphics[scale=0.5]{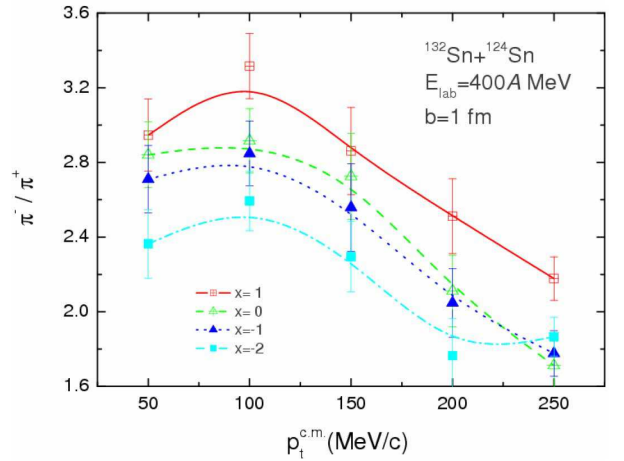}
\caption{(Color online) The $\protect\pi ^{-}/\protect\pi ^{+}$ ratio as a
function of transverse momentum. Taken from Ref. \protect\cite{LiBA05a}.}
\label{pionratio}
\end{figure}

At beam energy of about 400 MeV/nucleon, pion production becomes
non-negligible and may also carry interesting information about the
EOS of dense neutron-rich matter \cite{LiBA03,LiBA05a}. In
Fig.~\ref{pionyield}, we show the $\pi ^{-}$ and $\pi ^{+}$ yields
as functions of the $x$ parameter in the MDI interaction. It is seen
that when the $x$ parameter is changed from -2 to 1, the $\pi ^{-}$
multiplicity increases by about 20\%, although the $\pi ^{+}$
multiplicity remains about the same. The $\pi^-$ multiplicity is
thus more sensitively to the symmetry energy than that of $\pi^+$.
Also, the multiplicity of $\pi ^{-}$ is about 2 to 3 times that of
$\pi ^{+}$. This is because $\pi ^{-}$ mesons are mostly produced
from neutron-neutron collisions, which are more frequent in
collisions of neutron-rich nuclei. Since the high density region is
more neutron rich for the softer symmetry energy as a result of
isospin fractionation \cite{LiBA05a}, the $\pi ^{-}$ multiplicity is
thus more sensitive to the isospin asymmetry of the reaction system
and the symmetry energy. However, it is well known that the pion
yield is also sensitive to the symmetric part of the nuclear EOS,
and it is thus hard to extract reliable information about the
symmetry energy from the $\pi ^{-}$ yield alone. The $\pi ^{-}/\pi
^{+}$ ratio is, on the other hand, a better probe as this ratio is
sensitive only to the difference in the chemical potentials for
neutrons and protons \cite{Ber80}. As well demonstrated in
Fig.~\ref{pionratio}, the $\pi ^{-}/\pi ^{+}$ ratio is quite
sensitive to the symmetry energy, especially at low transverse
momenta, and can be used to probe the high density behavior of
nuclear symmetry energy $E_{\rm sym}(\rho )$.

\subsection{Double n/p and $\protect\pi ^{-}/\protect\pi ^{+}$ ratio}

Because of the facts that the symmetry potentials have opposite
signs for neutrons and protons and are also generally smaller
compared to the isoscalar potential at the same density, most
observables proposed so far for studying the density dependence of
the nuclear symmetry energy employ differences or ratios of isospin
multiplets of baryons, mirror nuclei and mesons, such as, the
neutron/proton ratio of emitted nucleons \cite{LiBA97a},
neutron-proton differential flow \cite{LiBA00}, neutron-proton
correlation function \cite{Che03a}, $t$/$^{3}$He
\cite{Che03b,Zha05}, $\pi ^{-}/\pi ^{+}$
\cite{LiBA02,Gai04,LiBA05a,LiQF05b}, $\Sigma ^{-}/\Sigma ^{+}$
\cite{LiQF05a} and $K^{0}/K^{+}$ ratios \cite{Fer05}, etc. Among
these observables, the ratio of emitted neutrons to protons has
probably the highest sensitivity to the symmetry energy as the
symmetry potential acts directly on nucleons and emitted nucleons
are also rather abundant in typical heavy-ion reactions. However, it
is very challenging to measure some of these observables, especially
those involving neutrons. The measurement of neutrons, particularly
the low energy ones, always suffers from low detection efficiencies
even for the most advanced neutron detectors. Therefore, observables
involving neutrons normally have large systematic errors. Moreover,
for essentially all of these observables, the Coulomb force on
charged particles plays an important role and sometimes competes
strongly with the symmetry potential. One has to disentangle
carefully effects of the symmetry potential from those due to the
Coulomb potential. It is thus very desirable to find experimental
observables which can reduce the influence of both the Coulomb force
and the systematic errors associated with neutrons. A possible
candidate for such an observable is the double ratios of emitted
neutrons and protons taken from two reaction systems using four
isotopes of the same element, namely, the neutron/proton ratio in
the neutron-rich system over that in the more symmetric system, as
recently proposed by Lynch {\it et al.} \cite{Lyn06}. They have
actually demonstrated the feasibility of measuring the double
neutron/proton ratios in central reactions of $^{124}$Sn$+^{124}$Sn
and $^{112}$Sn$+^{112}$Sn at a beam energy of $50$ MeV/nucleon at
the National Superconducting Cyclotron Laboratory \cite{Lyn06}.

Both the double neutron/proton ratio and the double $\pi ^{-}/\pi
^{+}$ ratio in $^{132}$Sn$+^{124}$Sn and $^{112}$Sn$+^{112}$Sn
reactions at $400$ MeV/nucleon have been studied in the IBUU model
in order to demonstrate the effect of symmetry energy at high
density. It was found that these ratios have about the same
sensitivity to the density dependence of symmetry energy as the
corresponding single ratio in the respective neutron-rich system
involved. Given the advantages of measuring the double ratios over
the single ones, the study of double ratios will be more useful for
further constraining the symmetry energy of neutron-rich matter.
Furthermore, the systematic errors associated with transport model
calculations are mostly related to the uncertainties in the
in-medium NN cross sections, techniques of treating collisions,
sizes of the lattices in calculating the phase space distributions,
techniques in handling the Pauli blocking, etc. Since the double
ratio is a relative observable from two similar reaction systems,
these systematic errors are expected to be reduced.

\begin{figure}[tbh]
\includegraphics[scale=0.85]{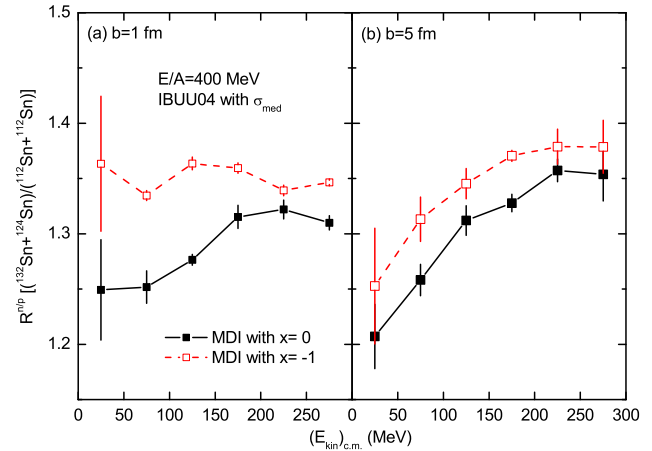}
\caption{(Color online) The double neutron/proton ratio of free
nucleons in the reaction of $^{132}$Sn $+^{124}$Sn and $^{112}$
Sn$+^{112}$Sn at $400$ MeV/nucleon and an impact parameter of $1$ fm
(left window) and $5$ fm (right window), respectively. Taken from
Ref. \protect\cite{LiBA06b}.} \label{DRatioNPE4h}
\end{figure}

In Fig.\ \ref{DRatioNPE4h}, we show the double neutron/proton
ratios from the reactions of $^{132}$Sn$+^{124}$Sn and
$^{112}$Sn$+^{112}$Sn at a beam energy of $400$ MeV/nucleon and an
impact parameter of $1$ fm (left window) and $5$ fm (right window)
predicted by the IBUU model using the MDI interaction with $x=0$
and $x=-1$, which are consistent with the symmetry energy used for
sub-saturation densities. At both impact parameters, effects of
the symmetry energy are about $5\%-10\% $ changing from the case
with $x=0$ to $x=-1$. One notices here that the low energy
nucleons have the largest sensitivity to the variation of the
symmetry energy for such high energy heavy-ion collisions. In
fact, the neutron/proton ratio of midrapidity nucleons which have
gone through the high density phase of the reaction are known to
be most sensitive to the symmetry energy \cite{LiBA05b}. Compared
to the results at the beam energy of $50$ MeV/nucleon
\cite{LiBA05b}, it is interesting to see a clear turnover in the
dependence of the double neutron/proton ratio on the $x $
parameter, namely the double ratio is lower at $50$ MeV/nucleon
but higher at $400$ MeV/nucleon with $x=-1 $ than that with $x=0$.
The maximum density reached at the beam energy of $50 $ and $400$
MeV/nucleon is about $1.2\rho _{0}$ and $2\rho _{0}$
\cite{LiBA05b}, respectively. The turnover clearly indicates that
the double neutron/proton ratio reflects closely the density
dependence of the symmetry energy as shown in Fig.~\ref{MDIEsym}.
This observation also indicates that systematic studies of the
double neutron/proton ratio over a broad beam energy range will be
important for mapping out the density dependence of the symmetry
energy.

\begin{figure}[tbh]
\includegraphics[scale=0.85]{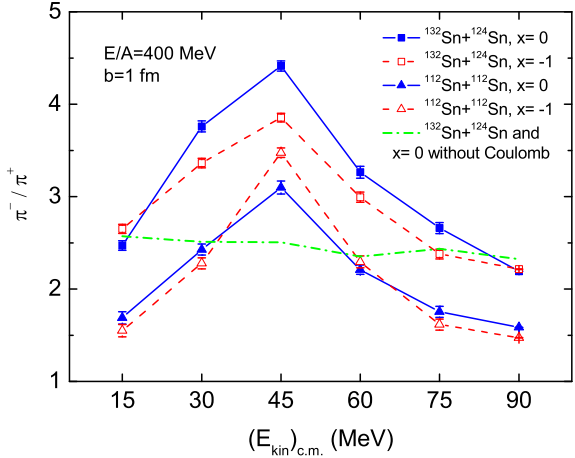}
\caption{(Color online) Kinetic energy distribution of the single
$\protect\pi ^{-}/\protect\pi ^{+}$ ratio for $^{132}$ Sn+$^{124}$
Sn and $^{112}$ Sn+$^{112}$ Sn at a beam energy of $400$ MeV/nucleon
and an impact parameter of $b=1$ fm with the stiff ($x=-1$) and soft
($x=0 $) symmetry energies. The dash-dotted line is the single
$\protect\pi ^{-}/\protect\pi ^{+}$ ratio obtained by turning off
the Coulomb potentials in the $^{132}$ Sn+$^{124}$ Sn reaction.
Taken from Ref. \protect\cite{Yon06a}.} \label{Rpion}
\end{figure}

As shown before from both the total yields and the momentum spectra,
the $\pi ^{-}/\pi ^{+}$ ratio is a promising probe of the symmetry
energy at high densities
\cite{LiBA02,LiBA05a,Gai04,LiQF05b,LiQF05a}. In Fig.~\ref{Rpion}, we
show again the kinetic energy distribution of the single $\pi
^{-}/\pi ^{+}$ ratio for the reactions of $^{132}$Sn+$^{124}$Sn and
$^{112}$Sn+$^{112}$Sn at a beam energy of $400$ MeV/nucleon and an
impact parameter of $b=1$ fm with the stiff ($x=-1$) or soft ($x=0$)
symmetry energy, obtained with $12000$ IBUU events for each
reaction. It is seen that the overall magnitude of $\pi ^{-}/\pi
^{+}$ ratio is larger for the neutron-rich system
$^{132}$Sn+$^{124}$Sn than for the neutron-deficient system
$^{112}$Sn+$^{112}$Sn as expected. Also, the soft symmetry energy
($x=0$) leads to a larger single $\pi ^{-}/\pi ^{+}$ ratio than the
stiff one ($x=-1$). This is mainly because the high density region
where most pions are produced are more neutron-rich with the softer
symmetry energy as a result of the isospin fractionation
\cite{LiBA02,LiBA05a}. Furthermore, the single $\pi ^{-}/\pi ^{+}$
ratio is more sensitive to the symmetry energy in the reaction
$^{132}$Sn+$^{124}$Sn than in the reaction $^{112}$Sn+$^{112}$Sn as
a result of the larger isospin asymmetry in the more neutron-rich
system.

Fig.~\ref{Rpion} shows that the single $\pi ^{-}/\pi ^{+}$ ratio
exhibits a peak at a pion kinetic energy of about $45$ MeV in all
cases considered here. The origin of this peak can be understood
from the single $\pi ^{-}/\pi ^{+}$ ratios in both reactions from
turning off the Coulomb potentials for all charged particles. As an
example, shown in Fig.\ \ref{Rpion} with the dash-dotted line is the
single $\pi ^{-}/\pi ^{+}$ ratio obtained by turning off the Coulomb
potentials in the $^{132}$Sn+$^{124}$Sn reaction. It is seen that
the single $\pi ^{-}/\pi ^{+}$ ratio now becomes approximately a
constant of about $2.4$. This value agrees with the predicted value
of $(5N^{2}+NZ)/(5Z^{2}+NZ)\approx (N/Z)^{2}\approx 2.43$, where $N$
and $Z$ are the total neutron and proton numbers in the participant
region, from the $\Delta $ resonance model \cite{Sto86a} for central
$^{132}$Sn+$^{124}$Sn collisions. This is not surprising as at $400$
MeV/nucleon, pions are almost exclusively produced via the $\Delta $
resonances \cite{Li91}. Comparison of calculated results with and
without the Coulomb potentials thus indicates clearly that the peak
observed in the single $\pi ^{-}/\pi ^{+}$ ratio is due to the
Coulomb effects. Although the Coulomb potential distorts the spectra
of pions, the effect of symmetry potential is still seen in the
resulting $\pi ^{-}/\pi ^{+}$, particularly near its peak value
where pions have relatively low kinetic energies. These pions are
produced in the high density nucleonic matter (about $2\rho _{0}$)
through the $\Delta $ resonances and experience many rescatterings
with nucleons at both high and low densities as well as the Coulomb
potential from protons at different densities. Since the soft
($x=0$) and stiff ($x=-1$) symmetry energies have slight difference
at low densities and the large difference appears at high densities
(about $2\rho _{0}$) as shown in Fig.~\ref{MDIEsym}, the observed
symmetry energy effects on the energy dependence of the $\pi
^{-}/\pi ^{+}$ ratio thus mainly reflect (though not completely)
information on the high density behavior of the symmetry energy
\cite{LiQF05b}.

\begin{figure}[tbh]
\includegraphics[scale=0.85]{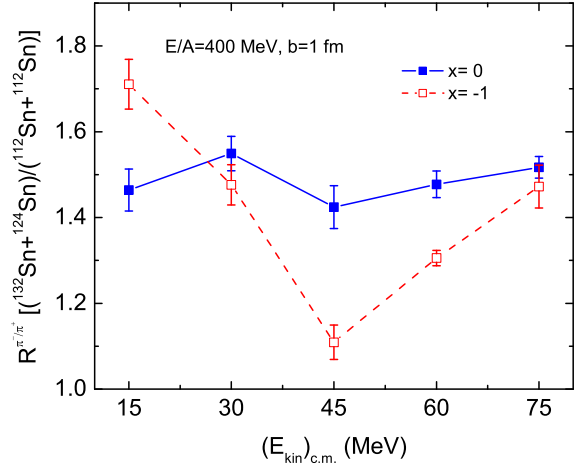}
\caption{(Color online) Kinetic energy dependence of the double
$\protect\pi ^{-}/\protect\pi ^{+}$ ratio of $^{132}$ Sn+$^{124}$ Sn
over $^{112}$ Sn+$^{112}$ Sn at a beam energy of $400$ MeV/nucleon
and an impact parameter $b=1 $ fm with the stiff ($x=-1$) and soft
($x=0$) symmetry energies. Taken from Ref. \protect\cite{Yon06a}.}
\label{DRpion}
\end{figure}

In order to reduce the systematic errors related to the symmetry
energy effect on the $\pi ^{-}/\pi ^{+}$ ratio, it is more useful to
study the double $\pi^{-}/\pi ^{+}$ ratio in the reactions of
$^{132}$Sn+$^{124}$Sn and $^{112}$Sn+$^{112}$Sn as for the double
proton/neutron ratio. Fig.~\ref{DRpion} shows the double $\pi
^{-}/\pi ^{+}$ ratio for these two reactions. It is seen that the
kinetic energy dependence of the double $\pi ^{-}/\pi ^{+}$ ratio is
rather different for the stiff ($x=-1 $) and soft ($x=0$) symmetry
energies. While the double $\pi ^{-}/\pi ^{+}$ ratio is quite flat
for $x=0$, it displays a concave structure for $x=-1$ around the
Coulomb peak. These different behaviors can be understood from the
corresponding single $\pi ^{-}/\pi ^{+}$ ratios in the two reactions
shown in Fig. \ref{Rpion}. It is reassuring to see that around the
Coulomb peak the double $\pi ^{-}/\pi ^{+}$ ratio is still sensitive
to the symmetry energy. Compared with the single $\pi ^{-}/\pi ^{+}$
ratio, the kinetic energy dependence of the double $\pi ^{-}/\pi
^{+}$ ratio becomes, however, weaker. This is because effects of the
Coulomb potentials are reduced in the double $\pi ^{-}/\pi ^{+}$
ratio. We note that the double $\pi ^{-}/\pi ^{+}$ ratio displays an
opposite symmetry energy dependence compared with the double $n/p$
ratio for free nucleons shown in Fig.~\ref{DRatioNPE4h}. This is
understandable since the soft symmetry energy leads to a more
neutron-rich dense matter in heavy-ion collisions induced by
neutron-rich nuclei and thus a smaller $n/p$ ratio for free nucleons
due to the charge conservation. On the other hand, more $\pi ^{-}$'s
would be produced due to more neutron-neutron inelastic scatterings
in the more neutron-rich matter.

\subsection{Double neutron-proton differential transverse flow}

The neutron-proton differential transverse flow defined in Eq.
(\ref{npdiffflow}) can be further expressed as \cite{LiBA00,
LiBA02,LiBA05a}
\begin{eqnarray}
F_{n-p}^{x}(y) &\equiv
&\frac{1}{N(y)}\sum_{i=1}^{N(y)}p_{i}^{x}(y)w_{i}
\notag \\
&=&\frac{N_{n}(y)}{N(y)}\langle p_{n}^{x}(y)\rangle
-\frac{N_{p}(y)}{N(y)}\langle p_{p}^{x}(y)\rangle  \label{npflow}
\end{eqnarray}
where $N(y)$, $N_{n}(y)$ and $N_{p}(y)$ are the numbers of free
nucleons, neutrons, and protons, respectively, at rapidity $y$;
$p_{i}^{x}(y)$ is the transverse momentum of a free nucleon at
rapidity $y$; $w_{i}=1$ $(-1)$ for neutrons (protons); and $\langle
p_{n}^{x}(y)\rangle $ and $\langle p_{p}^{x}(y)\rangle $ are,
respectively, the average transverse momenta of neutrons and protons
at rapidity $y$. Eq. (\ref{npflow}) shows that the neutron-proton
differential transverse flow depends not only on the proton and
neutron transverse momenta but also on their relative
multiplicities, i.e., the isospin fractionation. This can be more
clearly seen from the following two special cases. If neutrons and
protons have the same average transverse momentum in the reaction
plane but different multiplicities in each rapidity bin, i.e.,
$\langle p_{n}^{x}(y)\rangle =\langle p_{p}^{x}(y)\rangle =\langle
p^{x}(y)\rangle $, and $N_{n}(y)\neq N_{p}(y)$, then Eq.
(\ref{npflow}) is reduced to
\begin{equation}
F_{n-p}^{x}(y)=\frac{N_{n}(y)-N_{p}(y)}{N(y)}\langle p^{x}(y)\rangle =\delta
(y)\cdot \langle p^{x}(y)\rangle,
\end{equation}
reflecting the effects of isospin fractionation. On the other hand,
if neutrons and protons have the same multiplicity but different
average transverse momenta, i.e., $N_{n}(y)=N_{p}(y)$ but $\langle
p_{n}^{x}(y)\rangle \neq \langle p_{p}^{x}(y)\rangle $, then Eq.
(\ref{npflow}) is reduced to
\begin{equation}
F_{n-p}^{x}(y)=\frac{1}{2}(\langle p_{n}^{x}(y)\rangle -\langle
p_{p}^{x}(y)\rangle )
\end{equation}
and reflects directly the difference between the neutron and proton
transverse flows. Since a stiffer symmetry potential is expected to
lead to a higher isospin fractionation and also a larger transverse
flow for neutrons than for protons in heavy-ion collisions at higher
energies, the neutron-proton differential flow is thus a measure of
these two combined effects of the symmetry potentials on neutrons
and protons \cite{LiBA06b,Yon06a}.

\begin{figure}[tbh]
\includegraphics[width=0.5\textwidth]{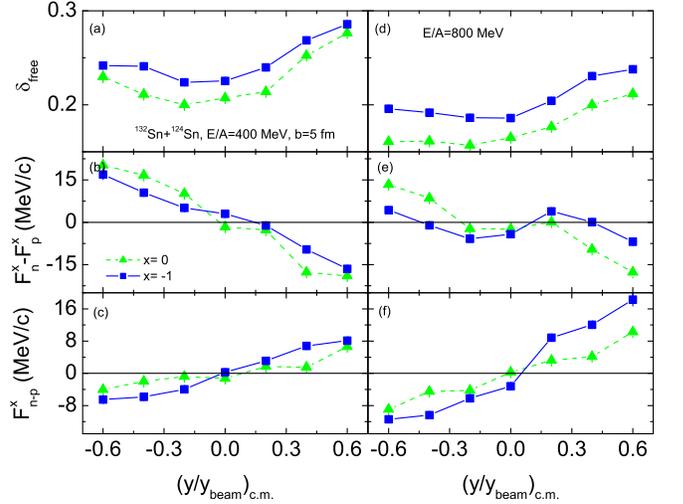}
\caption{(Color online) Rapidity distributions of the isospin
asymmetry of free nucleons (upper panels), the difference of the
average nucleon transverse flows (middle panels) and the
neutron-proton differential transverse flow (lower panels) from
$^{132}$Sn+$^{124}$Sn reaction at the incident beam energies of
$400$, $800$ MeV/nucleon and $b=5$ fm with two symmetry energies of
$x=0$ and $x=-1$. Taken from Ref. \protect\cite{Yon06b}. }
\label{isoflow1}
\end{figure}

Shown in Fig.~\ref{isoflow1} are the rapidity distributions of the
isospin asymmetry of free nucleons (upper panels), the difference
of the average nucleon transverse flows (middle panels) and the
neutron-proton differential transverse flow (lower panels) from
the $^{132}$Sn+$^{124}$Sn reaction at incident beam energies of
$400$ and $800$ MeV/nucleon and an impact parameter of $b=5$ fm
with the two symmetry energies of $x=0$ and $x=-1$. It is seen
that the stiffer symmetry energy ($x=-1 $) leads to a larger
isospin asymmetry of free nucleons (stronger isospin
fractionation) (upper panels) than the softer symmetry energy
($x=0$). As a result, the neutron-proton differential transverse
flow from the stiff symmetry energy ($x=-1$) (bottom panels) is
much larger than that from the soft symmetry energy ($x=0$)
compared to the difference between the average neutron and proton
transverse flows obtained from the two symmetry energies (middle
panels). Also, the negative (positive) values of the neutron and
proton average transverse flow $F_n^x-F_p^x$ at forward (backward)
rapidities as a result of the Coulomb potential effect on protons
is reversed for the neutron-proton differential flow
$F_{n-p}^{x}(y)$ after taking into account the effect due to
isospin fractionation.

From panels (c) and (f) of Fig.\ \ref{isoflow1}, it is seen that the
slope of the neutron-proton differential transverse flow around the
mid-rapidity obtained from the same symmetry energy is larger for
the higher incident beam energy. This is mainly because a denser
nuclear matter is formed at higher incident beam energy, leading to
a stronger symmetry potential and thus higher transverse momenta for
neutrons compared to protons. Although the net effect of the
symmetry potential on the neutron-proton differential transverse
flow at 800 MeV/nucleon is not much larger than that at $400$
MeV/nucleon, its magnitude is much larger and is thus easier to be
measured experimentally.

\begin{figure}[tbh]
\includegraphics[width=0.5\textwidth]{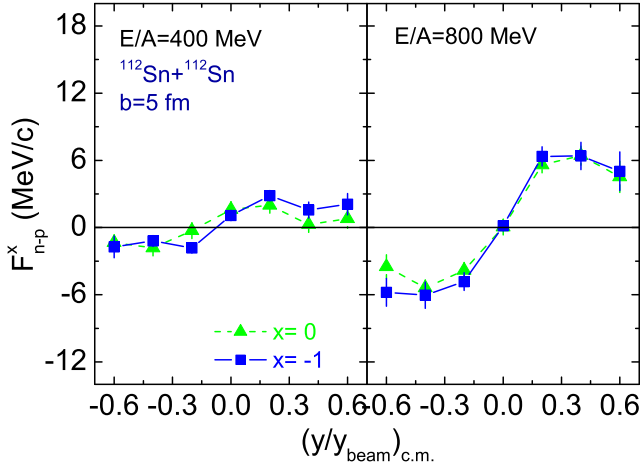}
\caption{(Color online) Same as panels (c) and (f) of Fig.\
\protect\ref{isoflow1} but for the reaction system of
$^{112}$Sn+$^{112}$Sn. Taken from Ref. \protect\cite{Yon06b}.}
\label{flow112}
\end{figure}

The systematic errors in the neutron-proton differential flow can be
reduced by studying its values in two similar reaction systems.
Besides the $^{132}$Sn+$^{124}$Sn reaction involving neutron-rich
nuclei, one can consider another less neutron-rich reaction system
$^{112}$Sn+$^{112} $Sn. Fig.\ \ref{flow112} shows the rapidity
distribution of the neutron-proton differential transverse flow in
the semi-central reaction of $^{112}$Sn+$^{112}$Sn at the same
incident beam energies of $400$ and $800$ MeV/nucleon. In comparison
with the $^{132}$Sn+$^{124}$Sn reaction, the slope of the
neutron-proton differential transverse flow around mid-rapidity and
effects of the symmetry energy are much smaller due to the smaller
isospin asymmetry in the reaction of $^{112}$Sn+$^{112}$Sn.

\begin{figure}[tbh]
\includegraphics[width=0.5\textwidth]{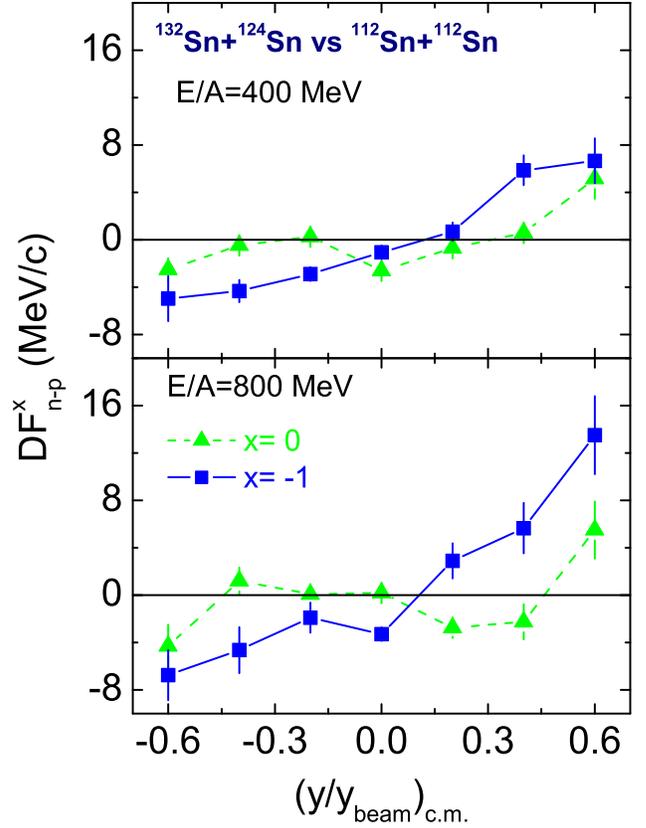}
\caption{(Color online) Rapidity distribution of the double
neutron-proton differential transverse flow in semi-central
reactions of Sn+Sn isotopes at incident beam energies of $400$ and
$800$ MeV/nucleon with two symmetry energies of $x=0$ and $x=-1$.
Taken from Ref. \protect\cite{Yon06b}. } \label{dflow}
\end{figure}

The double neutron-proton differential flow is defined as the
difference of the neutron-proton differential flows in the two
reaction systems of $^{132}$Sn+$^{124}$Sn and $^{112}$Sn+$^{112}$Sn.
Fig.\ \ref{dflow} shows the rapidity distribution of the double
neutron-proton differential transverse flow in the semi-central
reactions of Sn+Sn isotopes. At both incident beam energies of $400$
and $800$ MeV/nucleon, the double neutron-proton differential
transverse flow around mid-rapidity is essentially zero for the soft
symmetry energy of $x=0$. It displays, however, a finite slope with
respect to the rapidity for the stiffer symmetry energy of $x=-1$.
Moreover, the double neutron-proton differential transverse flow at
the higher incident energy exhibits a stronger symmetry energy
effect as expected. Since the double neutron-proton differential
transverse flow retains about the same symmetry energy effect as the
$^{132}$Sn+$^{124}$Sn reaction, it is less sensitive to the
systematic uncertainties in experiments \cite{Ram00}.

\begin{figure}[tbh]
\includegraphics[width=0.5\textwidth]{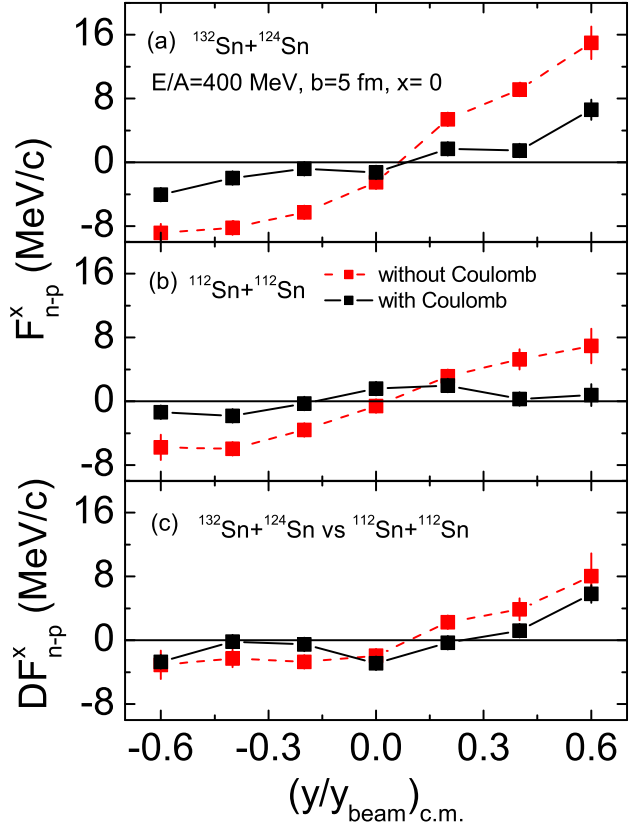}
\caption{(Color online) Coulomb effects on the neutron-proton
differential transverse flow (upper two panels) and the double
neutron-proton differential transverse flow (lowest panel) in the
semi-central reactions of Sn+Sn isotopes at the incident beam energy
of $400$ MeV/nucleon with the symmetry energy of $x=0$. Taken from
Ref. \protect\cite{Yon06b}.} \label{Coulomb}
\end{figure}

Also, the Coulomb effect, which competes strongly with the symmetry
potentials, is less important in the double neutron-proton
differential transverse flow than in the neutron-proton differential
transverse flow. This can be seen in Fig.\ \ref{Coulomb} which shows
the neutron-proton differential transverse flow (upper two panels)
and the double neutron-proton differential transverse flow (lowest
panel) in the semi-central reactions of Sn+Sn isotopes at the
incident beam energy of $400$ MeV/nucleon with the symmetry energy
of $x=0$ for the two cases of with and without the Coulomb
potential. From the upper two panels of Fig.\ \ref{Coulomb}, one
sees that the Coulomb effect reduces the strength of the
neutron-proton differential transverse flow as it makes more protons
unbound and to have large transverse momenta in the reaction-plane.
The Coulomb effect is, however, largely reduced in the double
neutron-proton differential transverse flow shown in the bottom
panel of Fig.\ \ref{Coulomb}.

\subsection{The n/p ratio of squeezed-out nucleons}

It is well known that in noncentral heavy-ion collisions nucleons
in the participant region is squeezed out of the reaction plane as
a result of the large density gradient in this direction and the
absence of spectator nucleons to block their emissions. These
nucleons can thus carry direct information about the high density
phase of the reaction and have been widely used in probing the EOS
of dense matter, particularly that of the symmetric nuclear
matter, see, e.g., Refs.
\cite{Sto86b,Ber88b,Cas90,Aic91,Rei97,Dan02a} for a review.

\begin{figure}[th]
\begin{center}
\includegraphics[width=0.5\textwidth]{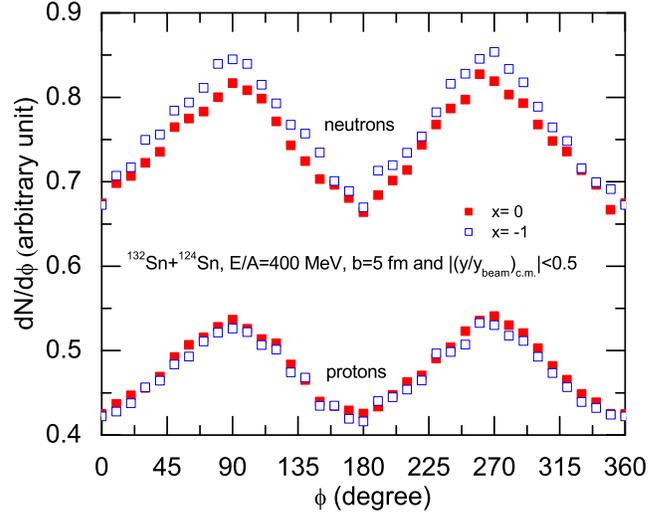}
\end{center}
\caption{(Color online) Azimuthal distribution of midrapidity nucleons
emitted in the reaction of $^{132}$Sn+$^{124}$Sn at an incident beam energy
of $400$ MeV/nucleon and an impact parameter of $b=5$ fm. Taken from Ref.
\protect\cite{Yon07}.}
\label{degree}
\end{figure}

The possibility of using the squeezed out nucleons to study the
high density behavior of the nuclear symmetry energy has been
studied recently using the IBUU04 model \cite{Yon07}. Shown in
Fig.\ \ref{degree} are the azimuthal distributions of free
nucleons in the midrapidity region ($|(y/y_{beam})_{c.m.}|<0.5$)
of the reaction $^{132}$Sn+$^{124}$Sn at an incident beam energy
of $400$ MeV/nucleon and an impact parameter of $b=5$ fm,
predicted by the IBUU model using the MDI interactions with $x=0$
and $-1$. A preferential emission of nucleons perpendicular to the
reaction plane is clearly observed for both neutrons and protons.
The sensitivity to the nuclear symmetry energy is mainly seen in
the squeezed-out neutrons, which experience a stronger repulsive
symmetry potential for a stiff ($x=-1$) than for a soft ($x=0$)
nuclear symmetry energy in the neutron-rich matter, in addition to
the strong nuclear isoscalar potential. For protons, their
azimuthal distributions is less sensitive to the stiffness of the
nuclear symmetry energy. This is due to the additional repulsive
Coulomb potential which works against the attractive symmetry
potential experienced by protons in the neutron-rich matter.
Although it is not easy to measure neutrons in experiments, both
the transverse flow and squeeze-out of neutrons together with
other charged particles were measured at both the BEVALAC
\cite{Htu99} and SIS/GSI \cite{Ven93,Lei93,Lam93}. These
experiments and the associated theoretical calculations, see,
e.g., Refs. \cite{Bas95,Lar00}, have, however, all focused on
extracting information about the EOS of symmetric nuclear matter
without paying much attention to the effects due to the nuclear
symmetry energy.

\begin{figure}[th]
\begin{center}
\includegraphics[width=0.5\textwidth]{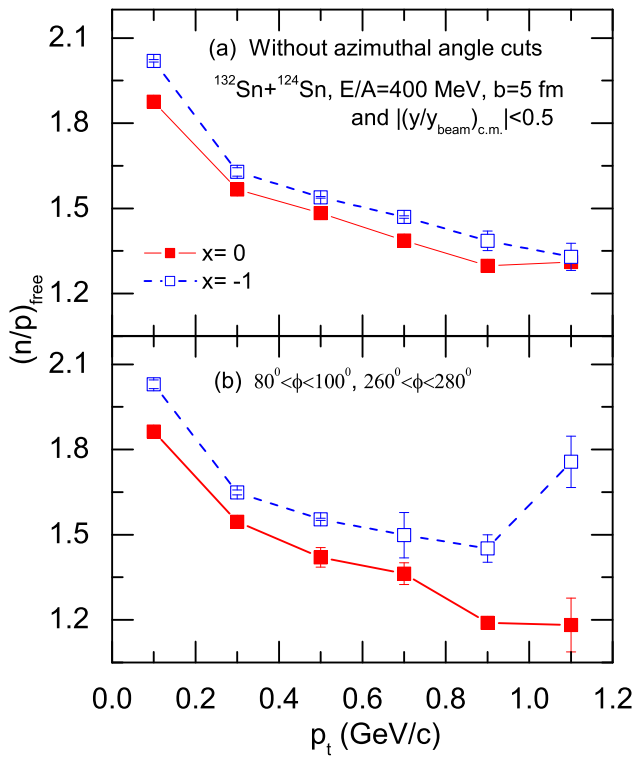}
\end{center}
\caption{(Color online) Transverse momentum distribution of the
ratio of midrapidity neutrons to protons emitted in the reaction
of $^{132}$Sn+$^{124} $Sn at the incident beam energy of $400$
MeV/nucleon and impact parameter of $b=5$ fm, with (lower window)
and without (upper window) an azimuthal angle cut of $80^{\circ
}<\protect\phi <100^{\circ }$ and $260^{\circ }<\protect\phi
<280^{\circ }$, which corresponds to squeezed-out nucleons emitted
in the direction perpendicular to the reaction plane. Taken from
Ref. \protect\cite{Yon07}.} \label{RnpSq}
\end{figure}

To reduce the effect due to uncertainties associated with the EOS
of symmetric nuclear matter, it is useful to consider the ratio of
squeezed-out neutrons to protons, particularly its transverse
momentum dependence. As shown in Refs.~\cite{LiBA98,LiBA97a}, the
n/p ratio is determined mostly by the density dependence of the
symmetry energy and almost not affected by the EOS of symmetric
nuclear matter. In Fig.\ \ref{RnpSq}, we show the transverse
momentum dependence of the neutron/proton (n/p) ratio of
midrapidity nucleons emitted in the reaction of $^{132}$Sn+$^{124}
$Sn at the incident beam energy of $400$ MeV/nucleon and impact
parameter of $b=5$ fm. For squeezed-out nucleons emitted in the
direction perpendicular to the reaction plane, which are obtained
by introducing an azimuthal angle cut of $80^{\circ }<\protect\phi
<100^{\circ }$ and $260^{\circ }<\protect\phi <280^{\circ }$, the
symmetry energy effect on the n/p ratio increases with the
increasing transverse momentum $p_{t}$ as shown in the lower
window. The effect can be as large as $40\%$ at a transverse
momentum of $1$ GeV/c. Since high $p_{t}$ particles most likely
come from the high density region in the early stage during
heavy-ion collisions, they are thus more sensitive to the high
density behavior of the symmetry energy. Without the cut on the
azimuthal angle, the n/p ratio of free nucleons in the midrapidity
region is much less sensitive to the symmetry energy in the whole
range of transverse momentum. is as shown in the upper window. It
is worth mentioning that the n/p ratio of free nucleons
perpendicular to the beam direction in the CMS frame in
$^{124}$Sn+$^{124}$Sn reactions at $50$ MeV/nucleon was recently
measured at the NSCL/MSU \cite{Fam06}. This measurement was useful
for studying the density dependence of the symmetry energy at
sub-normal densities. To investigate the symmetry energy at
supra-normal densities, similar measurements that allow the
construction of the reaction plane using a TPC
(Time-Projection-Chamber) and simultaneous detection of neutrons
together with charged particles at much higher energies are being
planned \cite{Bic07}. The results reviewed here provide strong
scientific motivations and support for such experimental efforts.
Compared to other potential probes, the n/p ratio of squeezed-out
nucleons is complementary but carries more direct information
about the symmetry energy at high densities. The sensitivity to
the high density behavior of the nuclear symmetry energy observed
in the n/p ratio of squeeze-out nucleons is probably the highest
found so far among all observables studied within the same
transport model.

\subsection{$K^{0}/K^{+}$ and $\Sigma ^{-}/\Sigma ^{+}$ ratios}

Since the proposal of Aichelin and Ko that the kaon yield in heavy
ion collisions at energies that are below the threshold for kaon
production in a nucleon-nucleon collision in free space may be a
sensitive probe of the EOS of nuclear matter at high densities
\cite{Aic85}, a lot of works have been done both theoretically and
experimentally on this problem
\cite{Cas90,Ko96,Ko97,Cas99,Kol05,Fuc06a}. Since the kaon is an
iso-doublet meson with the quark content of $d\overline{s}$ for
$K^{0}$ and $u\overline{s}$ for $K^{+}$, the $K^{0}/K^{+}$ ratio
provides a potentially good probe of the nuclear symmetry energy
as the $n/p$ and $\pi^-/\pi^+$ ratios, especially its high density
behavior as kaons are produced mainly from the high density region
during the early stage of the reaction and suffer negligible
absorption effects.

\begin{figure}[tbp]
\includegraphics[scale=1.0]{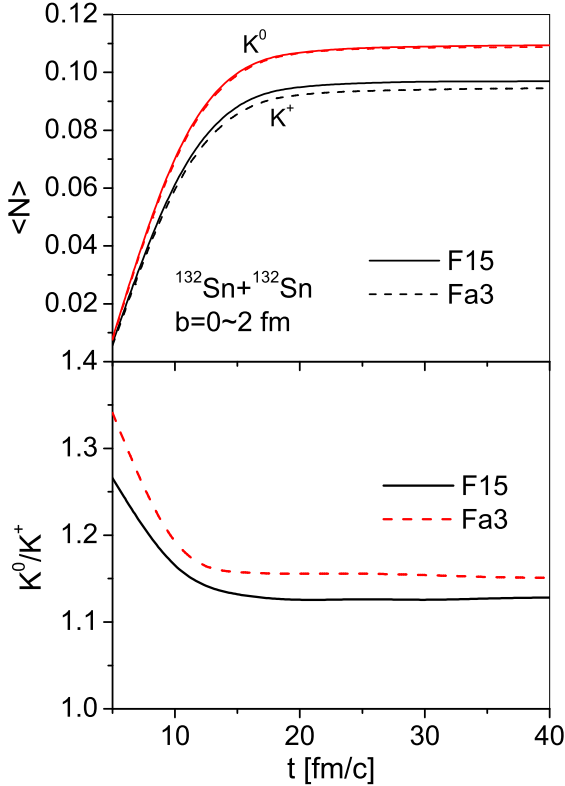}
\caption{(Color online) Top: Time evolution of the K$^{0}$ and
K$^{+}$ abundances for central $^{132}{\rm Sn}+^{132}{\rm Sn}$
collisions at a beam energy $1.5A $ GeV and with the symmetry
potentials F15 and Fa3. Bottom: the corresponding time evolution of
the $K^{0}/K^{+}$ ratios. Taken from Ref. \protect\cite{LiQF05c}.}
\label{RatioKaonLi}
\end{figure}

Using the UrQMD model (version 1.3), Li \textit{et al.} have
investigated the symmetry energy effects on the $K^{0}/K^{+}$ ratio
by studying $K^{0}$ and $K^{+}$ production from the central
$^{132}$Sn$+^{132}$Sn collisions at a beam energy $1.5A$ GeV with
two different forms of the symmetry energy, namely, the F15 and Fa3,
and the results are shown in Fig.\ \ref{RatioKaonLi} \cite{LiQF05c}.
Since the beam energy is close to the kaon production threshold,
which is about $1.58$ GeV for nucleon-nucleon collisions in free
space, the $K^{0}/K^{+}$ ratio displays only a small symmetry energy
effect. With decreasing beam energy, the symmetry energy effect
becomes larger. For example, the kaon yields from the reaction
$^{208}$Pb$+^{208}$Pb at $E_{\mathrm{b}}=0.8$ $A$ GeV and $b=7\sim
9\ \mathrm{fm}$, the $K^{0}/K^{+}$ ratio for the stiff F15 is about
$1.25$, whereas it is about $1.4$ for the soft Fa3. We note nuclear
in-medium effects on kaon production are neglected in the UrQMD
model simulations.

\begin{figure}[t]
\includegraphics[scale=0.35]{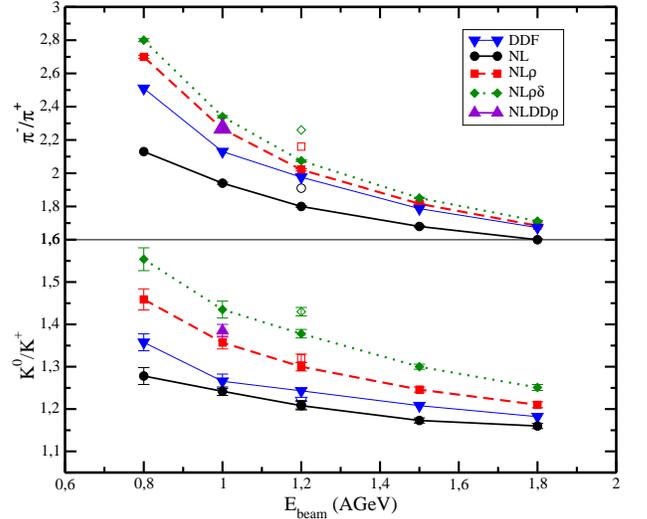}
\caption{ (Color online) $\protect\pi ^{-}/\protect\pi ^{+}$ (upper)
and $K^{+}/K^{0}$ (lower) ratios as a function of the incident
energy for central ($b=0$ fm impact parameter) Au+Au collisions with
the RBUU model . In addition, for $E_{beam}=1~AGeV$, $NL\protect\rho
$ results with a density dependent $\protect\rho $-coupling
(triangles) are also presented. The $open$ symbols at $1.2~AGeV$
show the corresponding results for a $^{132}Sn+^{124}Sn $ collision,
more neutron rich. Note the different scale for the $\protect\pi
^{-}/\protect\pi ^{+}$ ratios. Taken from Ref.
\protect\cite{Fer05}.} \label{RatioKaonFe}
\end{figure}

Above results were confirmed by Ferini et al. using a relativistic
hadronic transport model of Boltzmann-Uehling-Uhlenbeck type (RBUU)
with different forms of the symmetry energies in central ($b=0$ fm
impact parameter) Au+Au collisions \cite{Fer05}. Their results,
shown in Fig.\ \ref{RatioKaonFe}, indicate that at beam energies
below and around the kinematical threshold of kaon production, the
$K^{0}/K^{+}$ inclusive yield ratio is more sensitive to the
symmetry energy than the $\pi ^{-}/\pi ^{+}$, and subthreshold kaon
production thus could provide a promising tool to extract
information on the density dependence of the nuclear symmetry
energy.

\begin{figure}[th]
\includegraphics[width=8cm]{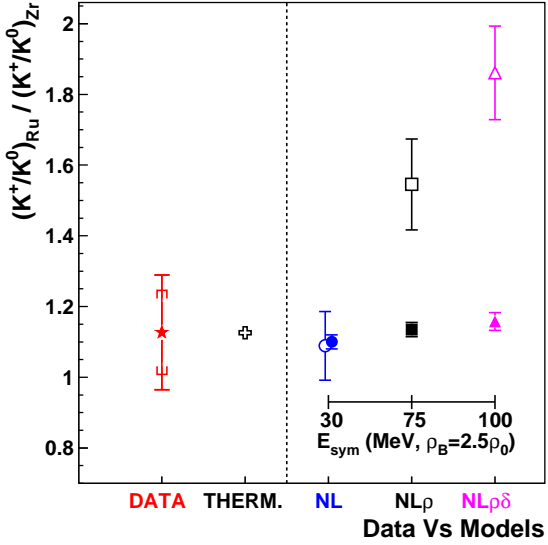}
\caption{(Color online) Experimental ratio
($K^{+}/K^{0}$)$_{Ru}$/($K^{+}/K^{0}$)$_{Zr}$ (star) and theoretical
predictions of the thermal model (cross) and the transport model
with 3 different assumptions on the symmetry energy: NL (circles),
NL$\protect\rho $ (squares) and NL$\protect\rho \protect\delta $
(triangles). The INM and HIC calculations are represented by open
and full symbols, respectively (see text for more details). The
statistic and systematic errors are represented by vertical bars and
brackets, respectively. Taken from Ref. \protect\cite{Lop07}.}
\label{DRatioKaon}
\end{figure}

Experimentally, the FOPI collaboration has reported recently the
results on $K^{+}$ and $K^{0}$ meson production in $_{44}^{96}$Ru +
$_{44}^{96}$Ru and $_{40}^{96}$Zr + $_{40}^{96}$Zr collisions at a
beam kinetic energy of $1.528$ $A$ GeV, measured with the FOPI
detector at GSI-Darmstadt \cite{Lop07}. The measured double ratio
($K^{+}/K^{0}$)$_{Ru}$/($K^{+}/K^{0}$)$_{Zr}$ is compared in Fig.\
\ref{DRatioKaon} to the predictions of a thermal model and the RBUU
transport model using two different collision scenarios and under
different assumptions on the stiffness of the symmetry energy. From
Fig.\ \ref{DRatioKaon}, one can see a good agreement with the
thermal model prediction and the assumption of a soft symmetry
energy for infinite nuclear matter while more realistic transport
simulations of the collisions show a similar agreement with the data
but exhibit a reduced sensitivity to the symmetry term. We note that
in the present RBUU calculations, the isospin dependence of the
$K^{+}$- and $K^{0}$-nucleon potentials in the asymmetric nuclear
medium has been neglected. Recently, Mishra \textit{et al.} studied
the isospin dependent kaon and antikaon optical potentials in dense
hadronic matter using a chiral SU(3) model and their results
indicate that the density dependence of the isospin asymmetry is
appreciable for the kaon and antikaon optical potentials. On the
other hand, subthreshold kaon production in heavy-ion collisions
depends on some detailed implementations of the transport model
\cite{Kol05,Fuc06a}. Therefore, extracting useful information on the
high density behavior of the nuclear symmetry energy from
subthreshold kaon production in heavy-ion collisions induced by
neutron-rich nuclei needs further studies from both experimental and
theoretical sides.

\begin{figure}[tbp]
\includegraphics[scale=0.75]{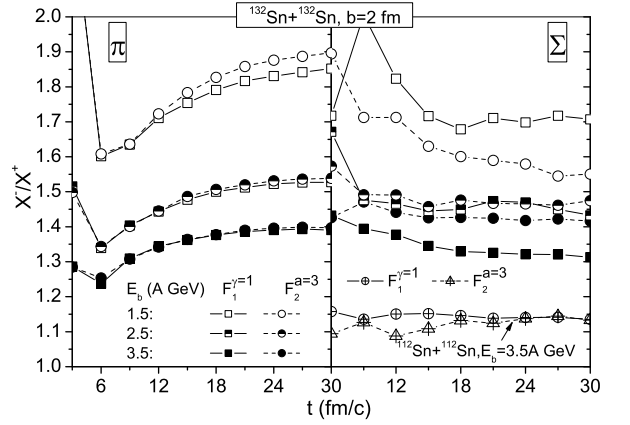}
\caption{The ratios $\protect\pi ^{-}/\protect\pi ^{+}$ (left) and
$\Sigma ^{-}/\Sigma ^{+}$ (right) for the collisions
$^{132}Sn+^{132}Sn$ ($E_{b}=1.5A $, $2.5A$, and $3.5A$ GeV; $b=2$
fm) and $^{112}Sn+^{112}Sn$ ($E_{b}=3.5A$ GeV; $b=2$ fm), calculated
with the different symmetry potentials $F_{1}^{\protect\gamma =1}$
and $F_{2}^{a=3}$. Taken from Ref. \protect\cite{LiQF05a}.}
\label{RatioSigma}
\end{figure}

Besides the $K^{0}/K^{+}$ ratio, the $\Sigma ^{-}/\Sigma ^{+}$ ratio
has also been proposed as a probe of the high density behavior of
the nuclear symmetry energy based on the UrQMD model (version 1.3)
calculations \cite{LiQF05a}. Shown in Fig.\ \ref{RatioSigma} is the
time evolution of the $\pi ^{-}/\pi ^{+}$ ratios (left-hand side)
and the $\Sigma ^{-}/\Sigma ^{+}$ ratios (right-hand side)
calculated with a stiff symmetry energy $F_{1}^{\gamma =1}$ and a
soft symmetry energy $F_{2}^{a=3}$ for the reaction
$^{132}$Sn$+^{132}$Sn at $E_{beam}=1.5A$, $2.5A$, $3.5A$ GeV and
$b=2$ fm, and $^{112}$Sn$+^{112}$Sn at $E_{b}=3.5A$ GeV and $b=2$
fm. It is seen that the $\Sigma ^{-}/\Sigma ^{+}$ ratio is sensitive
to the density dependence of the symmetry energy for neutron-rich
$^{132}$Sn$+^{132}$Sn collisions, but insensitive to that for the
nearly symmetric $^{112}$Sn$+^{112}$Sn collisions. For
$^{132}$Sn$+^{132}$Sn at $E_{b}=1.5A$ GeV, the $\Sigma ^{-}/\Sigma
^{+}$ ratio calculated with the stiff symmetry energy
($F_{1}^{\gamma =1}$) is higher than the one with the soft symmetry
energy ($F_{2}^{a=3}$). As the beam energy increases, the $\Sigma
^{-}/\Sigma ^{+}$ ratio falls and the difference between the $\Sigma
^{-}/\Sigma ^{+}$ ratios calculated with $F_{1}^{\gamma =1}$ and
$F_{2}^{a=3}$ reduces strongly. As the beam energy increases
further, at $E_{b}=3.5A$ GeV the $\Sigma ^{-}/\Sigma ^{+}$ ratio
falls further but the difference between the $\Sigma ^{-}/\Sigma
^{+}$ ratios calculated with $F_{1}^{\gamma =1}$ and $F_{2}^{a=3} $
appears again, the $\Sigma ^{-}/\Sigma ^{+}$ ratio with soft
symmetry energy now becoming higher than that with the stiff one.
For pions, the results indicate that the ratio $\pi ^{-}/\pi ^{+}$
at high energies (as in the case with $E_{b}=3.5A$ GeV) becomes
insensitive to the symmetry energy. The difference between the
$\Sigma ^{-}/\Sigma ^{+}$ ratio and the $\pi ^{-}/\pi ^{+}$ ratio
can be understood from the fact that, like nucleons, $\Sigma ^{\pm
}$ hyperons are under the influence of the mean field produced by
the surrounding nucleons, as soon as they are produced. The symmetry
potential of hyperons thus play an important dynamic role and
results in a strong effect on the ratio of the negatively to
positively charged $\Sigma $ hyperons.

\section{Summary and outlook}

\label{summary}

Heavy-ion collisions induced by neutron-rich nuclei provides a
unique opportunity for investigating the properties of the isospin
asymmetric nuclear matter, especially the density dependence of
the nuclear symmetry energy. To extract useful information from
these collisions, transport models have been found to be extremely
useful. Applications of these models have helped us understand not
only the isospin dependence of the in-medium nuclear effective
interactions but also that of the thermal, mechanical and
transport properties of asymmetric nuclear matter. These
information, particularly the density dependence of the nuclear
symmetry energy, are very important for both nuclear physics and
astrophysics. Significant progress has been made in recent years
in determining the density dependence of the nuclear symmetry
energy. Based on transport model calculations, a number of
sensitive probes of the symmetry energy have been identified. In
particular, the momentum dependence in both the isoscalar and
isovector parts of the nuclear potential was found to play an
important role in extracting accurately the density dependence of
the symmetry energy. From comparison of results from the transport
model with recent experimental data on isospin diffusion from
NSCL/MSU, a symmetry energy of $E_{sym}(\rho )\approx 31.6(\rho
/\rho _{0})^{\gamma }$ with $\gamma =0.69-1.05$ at subnormal
densities, which corresponds to the isospin and momentum dependent
MDI interaction with $x=0$ and $-1$, has been extracted. This
conclusion is consistent with those extracted from studying other
observables such as the isoscaling data and the neutron-skin
thickness in $^{208}$Pb.

\begin{figure}[th]
\begin{center}
\includegraphics[scale=0.90]{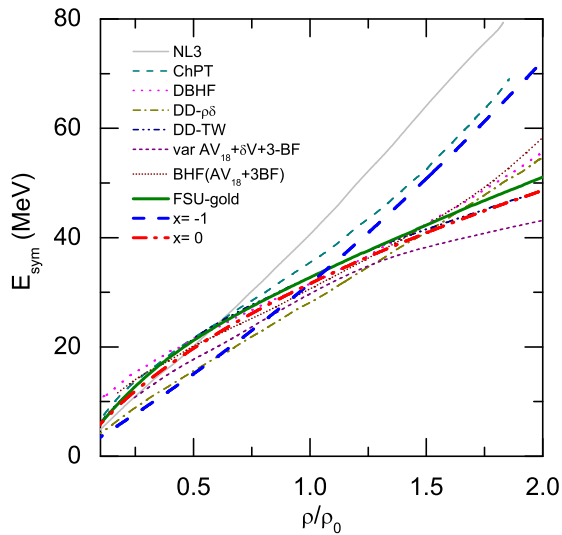}
\end{center}
\caption{(Color online) Density dependence of the nuclear symmetry
energy using the MDI interaction with $x=0$ and $x=-1$ and other
many-body theories predictions (data are taken from
\protect\cite{Fuc06b,Zuo02,Tod05}). Taken from Ref.
\protect\cite{Yon07}.} \label{EsymCurrent}
\end{figure}

Although considerable progress has been made in determining the
density dependence of the nuclear symmetry energy at sub-normal
densities, probing the high density behavior of the nuclear
symmetry energy remains a major challenge. Our current knowledge
on the density dependence of the nuclear symmetry energy can be
seen from Fig. \ref{EsymCurrent}, where predictions from several
typical theoretical model \cite{Die03,Fuc06b,Zuo02} are compared
with the phenomenological constraints that have been obtained from
model analysis of experimental data. The constraints labeled $x=0$
and $x=-1$ were extracted from studying the isospin diffusion in
the reaction of $^{124}$Sn +$^{112}$Sn at $E_{\rm beam}=50$ AMeV
within an isospin and momentum dependent transport model
\cite{Shi03,Tsa04,Che05a,LiBA05c}. For this particular reaction,
the maximum density reached is about $1.2\rho _{0}$. Moreover, it
was shown that the neutron-skin thickness in $^{208}$Pb calculated
within the Hartree-Fock approach using the same underlying Skyrme
interactions as the ones labeled $x=0$ and $x=-1$ is consistent
with the available experimental data \cite{Ste05b,LiBA06a,Che05b}.
The symmetry energy labeled as FSU-Gold was calculated within a
RMF model using an accurately calibrated parameter set such that
it reproduces both the giant monopole resonance in $^{90}$Zr and
$^{208}$Pb as well as the isovector giant dipole resonance of
$^{208}$Pb \cite{Tod05}. We note that the constraint obtained from
the isoscaling analysis is also consistent with the FSU-Gold and
the $x=0$ case \cite{She07}. At present, these results represent
the best phenomenological constraints on the nuclear symmetry
energy at sub-normal densities.

Although all predicted nuclear symmetry energies shown in Fig.
\ref{EsymCurrent} are close to the existing constraints at low
densities, they diverge widely at supra-normal densities,
including those from the MDI interaction with $x=-1$ and that with
$x=0$ as well as the FSU-Gold. Since there are currently no
experimental constraints on the high density behavior of the
nuclear symmetry energy, more work is thus needed. In particular,
experimental data including neutrons from reactions with
neutron-rich beams in a broad energy range will be useful for
studying the behavior of the symmetry energy at high densities. We
have reviewed recent theoretical progress in identifying the
observables in heavy-ion collisions induced by high energy
radioactive nuclei that are sensitive to the high density behavior
of the nuclear symmetry energy. A plethora of potentially
sensitive probes have been found, and they include the $\pi
^{-}/\pi ^{+}$ ratio, isospin fractionation, n-p differential
flow, double n/p and $\pi ^{-}/\pi ^{+}$ ratio, double n-p
differential transverse flow as well as the $K^{0}/K^{+}$ and
$\Sigma ^{-}/\Sigma ^{+}$ ratios. Studying these observable in
future experiments at high energy radioactive beam facilities is
expected to lead to significant constraints on the behavior of the
symmetry energy at supra-normal densities. Since transport models
are essential in extracting these constraints from the
experimental data, the continuous development of a practically
implementable quantum transport theory for nuclear reactions
induced by radioactive beams is important but poses a great
challenge.

\begin{acknowledgments}
This work was supported in part by the National Natural Science
Foundation of China under Grant Nos. 10575071, and 10675082, MOE of
China under project NCET-05-0392, Shanghai Rising-Star Program under
Grant No. 06QA14024, the SRF for ROCS, SEM of China, the US National
Science Foundation under Grant Nos. PHY-0652548 and PHY-0456890,
PHY-0457265, and the Welch Foundation under Grant No. A-1358.
\end{acknowledgments}


\begin{thebibliography}{999}
\bibitem{LiBA98} B.A. Li, C.M. Ko, and W. Bauer, topical review, Int. Jour.
Mod. Phys. E 7, 147 (1998).

\bibitem{LiBA01b} Isospin Physics in Heavy-Ion Collisions at Intermediate
Energies, Eds. Bao-An Li and W. Udo Schr\"{o}der (Nova Science Publishers,
Inc, New York, 2001).

\bibitem{Dan02a} P. Danielewicz, R. Lacey, and W.G. Lynch, Science 298, 1592
(2002).

\bibitem{Lat04} J.M. Lattimer and M. Prakash, Science 304, 536 (2004).

\bibitem{Bar05} V. Baran, M. Colonna, V. Greco, and M. Di Toro, Phys. Rep.
410, 335 (2005).

\bibitem{Ste05a} A.W. Steiner, M. Prakash, J.M. Lattimer, and P.J. Ellis,
Phys. Rep. 411, 325 (2005).

\bibitem{Mey66} W.D. Myers and W.J. Swiatecki, Nucl. Phys. A81, 1 (1966).

\bibitem{Pom03} K. Pomorski and J. Dudek, Phys. Rev. C 67, 044316 (2003).

\bibitem{Bom01} I. Bombaci, in \cite{LiBA01b}, p.35.

\bibitem{Die03} A.E.L. Dieperink, Y. Dewulf, D. Van Neck, M. Waroquier, and
V. Rodin, Phys. Rev. C 68, 064307 (2003).

\bibitem{Shl93} S. Shlomo and D. H. Youngblood, Phys. Rev. C 47, 529 (1993).

\bibitem{You99} D.H. Youngblood et al., 
Phys. Rev. Lett. 82, 691 (1999).

\bibitem{Bru67} K.A. Brueckner, S.A. Coon and J. Dabrowski, Phys. Rev.
\textbf{168}, 1184 (1967).

\bibitem{Sie70} P.J. Siemens, Nucl. Phys. \textbf{A141}, 225 (1970).

\bibitem{Sjo74} O. Sj\"oberg, Nucl. Phys. \textbf{A222}, 161 (1974).

\bibitem{Cug87} J. Cugnon, P. Deneye and A. Lejeune, Z. Phys. A \textbf{328},
409 (1987).

\bibitem{Bom91} I. Bombaci and U. Lombardo, Phys. Rev. C44, 1892 (1991).

\bibitem{Zuo02} W. Zuo, A. Lejeune, U. Lombardo, J. F. Mathiot, Eur. Phys.
J. A \textbf{14} (2002) 469.

\bibitem{Mut87} H. M\"uther, M. Prakash and T.L. Ainsworth, Phys. Lett.
\textbf{B199}, 469 (1987).

\bibitem{Har87} B. ter Haar and R. Malfliet, Phys. Rep. \textbf{149}, 207
(1987).

\bibitem{Sum92} K. Sumiyoshi, H. Toki and R. Brockmann, Phys. Lett. \textbf{B276},
393 (1992).

\bibitem{Hub93} H. Huber, F. Weber and M.K. Weigel, Phys. Lett. \textbf{B317},
485 (1993); Phys. Rev. C\textbf{50}, R1287 (1994).

\bibitem{Fuc04} E.N.E. van Dalen, C. Fuchs and A. Faessler, Nucl. Phys.
A741, 227 (2004); Phys. Rev. Lett. 95, 022302 (2005).

\bibitem{Ma04} Z.Y. Ma, J. Rong, B.Q. Chen, Z.Y. Zhu and H.Q. Song, Phys.
Lett. B604, 170 (2004).

\bibitem{Sam05a} F. Sammarruca, W. Barredo and P. Krastev, Phys. Rev. C71,
064306 (2005).

\bibitem{Mut00} H. M\"uther and A. Polls, Prog. Part. Nucl. Phys. \textbf{45},
243 (2000).

\bibitem{Dew02} Y. Dewulf, D. Van Neck, and M. Waroquier, Phys. Rev. C
\textbf{65}, 054316 (2002).

\bibitem{Car03} J. Carlson, J. Morales, Jr., V. R. Pandharipande, and D. G.
Ravenhall, Phys. Rev. C \textbf{68}, 025802 (2003).

\bibitem{Dic04} W.H. Dickhoff and C. Barbieri, Prog. Part. Nucl. Phys.
\textbf{52}, 377 (2004).

\bibitem{Fri81} B. Friedman and V.R. Pandharipande, Nucl. Phys. \textbf{A361},
502 (1981).

\bibitem{Lag81} I.E. Lagaris and V.R. Pandharipande, Nucl. Phys. \textbf{A369},
470 (1981)

\bibitem{Wir88a} R.B. Wiringa, V. Fiks and A. Fabrocini, Phys. Rev. C\textbf{38},
1010 (1988).

\bibitem{Akm98} A. Akmal, V. R. Pandharipande, and D. G. Ravenhall, Phys.
Rev. C\textbf{58}, 1804 (1998).

\bibitem{Muk07} A. Mukherjee and V. R. Pandharipande, Phys. Rev. C\textbf{75},
035802 (2007).

\bibitem{Ser97} B.D. Serot and J.D. Walecka, Int. Jour. Mod. Phys. E 6, 515
(1997).

\bibitem{Fur04} R.J. Furnstahl, Lect. Notes Phys. 641, 1 (2004).

\bibitem{Pra87} M. Prakash and T.L. Ainsworth, Phys. Rev. C \textbf{36}, 346
(1987).

\bibitem{Lut00} M. Lutz, B. Friman, and Ch. Appel, Phys. Lett. \textbf{B474},
7 (2000).

\bibitem{Fin04} P. Finelli, N. Kaiser, D. Vretenar and W. Weise, Nucl. Phys.
\textbf{A435}, 449 (2004).

\bibitem{Vre04} D. Vretenar and W. Weise, Lect. Notes Phys. 641, 65 (2004).

\bibitem{Fri05} S. Fritsch, N. Kaiser and W. Weise, Nucl. Phys. \textbf{A750},
259 (2005).

\bibitem{Fin06} P. Finelli, N. Kaiser, D. Vretenar and W. Weise, Nucl. Phys.
\textbf{A770}, 1 (2006).

\bibitem{Ser86} B.D. Serot and J.D. Walecka, Adv. Nucl. Phys. \textbf{16}, 1
(1986).

\bibitem{Chi77} S.A. Chin, Ann. Phys. (N.Y.), \textbf{108}, 301 (1977).

\bibitem{Hor87} C.J. Horowitz and B.D. Serot, Nucl. Phys. \textbf{A464}, 613
(1987); B.D. Serot and H. Uechi, Ann. Phys. (N.Y.) \textbf{179}, 272 (1987).

\bibitem{Gle82} N.K. Glendenning, Phys. Lett. \textbf{B114}, 392 (1982).

\bibitem{Hir91} D. Hirata \textit{et al.}, Phys. Rev. C\textbf{44}, 1467
(1991).

\bibitem{Sug94} Y. Sugahara and H. Toki, Nucl. Phys. \textbf{A579}, 557
(1994).

\bibitem{Rei89} P.-G. Reinhard, Rep. Prog. Phys. \textbf{52}, 439 (1989).

\bibitem{Rin96} P. Ring, Prog. Part. Nucl. Phys. \textbf{37}, 193 (1996).

\bibitem{Mil74} L.D. Miller, Phys. Rev. C\textbf{9}, 537 (1974).

\bibitem{Bro78} R. Brockmann, Phys. Rev. C\textbf{18}, 1510 (1978).

\bibitem{Jam81} M. Jaminon, C. Mahaux and P. Rochus, Nucl. Phys. \textbf{A365},
371 (1981).

\bibitem{Hor83} C.J. Horowitz and B.D. Serot, Nucl. Phys. \textbf{A399}, 529
(1983).

\bibitem{Bou87} A. Bouyssy, J.-F. Mathiot, N. Van Giai, and S. Marcos, Phys.
Rev. C\textbf{36}, 380 (1987).

\bibitem{Lop88} M. Lopez-Quelle et al.,
Nucl. Phys. A483, 479 (1988).

\bibitem{Ber93} P. Bernardos, V. N. Fomenko, Nguyen Van Giai, M. L. Quelle,
S. Marcos, R. Niembro, and L. N. Savushkin, Phys. Rev. C\textbf{48}, 2665
(1993).

\bibitem{Wer94} T.R. Werner \textit{et al.}, Phys. Lett. \textbf{B333}, 303
(1994).

\bibitem{Kho96} Dao T. Khoa, W. Von Oertzen and A.A. Ogloblin, Nucl. Phys.
\textbf{A602}, 98 (1996).

\bibitem{Vau72} D. Vautherin and D. M. Brink, Phys. Rev. C\textbf{5}, 626
(1972).

\bibitem{Bra85} M. Brack, C. Guet and H. -B. Hakansson, Phys. Rep.
\textbf{123}, 275 (1985).

\bibitem{Sto07} J.R. Stone and P.-G. Reinhard, Prog. Part. Nucl. Phys.
\textbf{58}, 587 (2007).

\bibitem{Kol85} K. Kolehmainen \textit{et al.}, Nucl. Phys. \textbf{A439},
535 (1985); J. Treiner \textit{et al.}, Ann. Phys. (N.Y.), \textbf{170}, 406
(1986).

\bibitem{Ban90} D. Bandyopadhyay, C. Samanta, S.K. Samaddar and J.N. De,
Nucl. Phys. \textbf{A511}, 1 (1990).

\bibitem{LiZH06} Z.H. Li, U. Lombardo, H.-J. Schulze, W. Zuo, L.W. Chen, and
H.R. Ma, Phys. Rev. C\textbf{74}, 047304 (2006).

\bibitem{LiBA95} B.A. Li and S.J. Yennello, Phys. Rev. C 52, 1746(R) (1995).

\bibitem{LiBA96} B.A. Li, Z.Z. Ren, C.M. Ko, and S.J. Yennello, Phys. Rev.
Lett. 76, 4492 (1996).

\bibitem{LiBA97a} B.A. Li, C.M. Ko, and Z.Z. Ren, Phys. Rev. Lett. 78, 1644
(1997).

\bibitem{LiBA97b} B.A. Li and C.M. Ko, Nucl. Phys. A618, 498 (1997).

\bibitem{Che97} L.W. Chen, L.X. Ge, X.D. Zhang, and F.S. Zhang, J. Phys. G
23, 211 (1997).

\bibitem{Che98} L.W. Chen, F.S. Zhang and G.M. Jin, Phys. Rev. C 58, 2283
(1998).

\bibitem{Bar98} V. Baran, M. Colonna, M. Di Toro, and A.B. Larionov, Nucl.
Phys. A632, 287 (1998).

\bibitem{Che99a} L.W. Chen, F.S. Zhang, G.M. Jin and Z.Y. Zhu, Phys. Lett.
\textbf{B459}, 21 (1999).

\bibitem{Zha99} F.S. Zhang, L.W. Chen, Z.Y. Ming and Z.Y. Zhu, Phys. Rev. C
\textbf{60}, 064604 (1999).

\bibitem{Xu00} H.S. Xu et al., Phys. Rev. Lett. 85, 716 (2000).

\bibitem{Che00} L.W. Chen, F.S. Zhang, and Z.Y. Zhu, Phys. Rev. C \textbf{61},
067601 (2000).

\bibitem{Zha00} F.S. Zhang, L.W. Chen, W.F. Li and Z.Y. Zhu, Eur. Phys. J.
\textbf{A9}, 149 (2000).

\bibitem{Tan01a} W.P. Tan et al., Phys. Rev. C 64, 051901(R) (2001).

\bibitem{Bar02} V. Baran, M. Colonna, M. Di Toro, V. Greco, and M.
Zielinska-Pfab\'{e}, and H.H. Wolter, Nucl. Phys. A703, 603 (2002).

\bibitem{Tsa01} M.B. Tsang et al., Phys. Rev. Lett. 86, 5023 (2001).

\bibitem{LiBA01a} B.A. Li, A.T. Sustich, and B. Zhang, Phys. Rev. C 64,
054604 (2001).

\bibitem{LiBA00} B.A. Li, Phys. Rev. Lett. 85, 4221 (2000).

\bibitem{Tan01b} Special issue on Radioactive Nuclear Beams, Edited by I.
Tanihata [Nucl. Phys. A693, (2001)].

\bibitem{LiBA02} B.A. Li, Phys. Rev. Lett. 88, 192701 (2002); Nucl. Phys.
A708, 365 (2002).

\bibitem{Che03a} L.W. Chen, V. Greco, C.M. Ko, and B.A. Li, Phys. Rev. Lett.
90, 162701 (2003); Phys. Rev. C 68, 014605 (2003)

\bibitem{Che03b} L.W. Chen, C.M. Ko, and B.A. Li, ibid C 68, 017601 (2003);
Nucl. Phys. A729, 809 (2003).

\bibitem{Ono03} A. Ono, P. Danielewicz, W.A. Friedman, W.G. Lynch, and M.B.
Tsang, Phys. Rev. C 68, 051601 (R) (2003).

\bibitem{Liu03} J.Y. Liu, W.J. Guo, Y.Z. Xing, and H. Liu, Nucl. Phys. A726,
123 (2003).

\bibitem{Che04} L.W. Chen, C.M. Ko, and B.A. Li, Phys. Rev. C 69, 054606
(2004).

\bibitem{LiBA04a} B.A. Li, C. B. Das, S. Das Gupta, and C. Gale, Phys. Rev.
C 69, 011603 (R) (2004); Nucl. Phys. A735, 563 (2004).

\bibitem{Shi03} L. Shi and P. Danielewicz, Phys. Rev. C 68, 064604 (2003).

\bibitem{LiBA04b} B.A. Li, Phys. Rev. C 69, 034614 (2004).

\bibitem{Riz04} J. Rizzo, M. Colonna, M. Di Toro, and V. Greco, Nucl. Phys.
A732, 202 (2004).

\bibitem{LiBA05a} B.A. Li, G.C. Yong and W. Zuo, Phys. Rev. C 71, 014608
(2005).

\bibitem{LiBA05b} B.A. Li, G.C. Yong and W. Zuo, Phys. Rev. C 71, 044604
(2005).

\bibitem{Zha05} Y. Zhang and Z. Li, Phys. Rev. C 71, 024604 (2005).

\bibitem{LiQF05a} Q. Li, Z. Li, E. Zhao, and R.K. Gupta, Phys. Rev. C 71,
054907 (2005).

\bibitem{Tia05} W.D. Tian et al., Chin. Phys. Lett. 22, 306 (2005).

\bibitem{Mul95} H. Muller and B. Serot, Phys. Rev. C 52, 2072 (1995).

\bibitem{Gai04} T. Gaitanos, M. Di Toro, S. Type, V. Baran, C. Fuchs, V.
Greco, H.H. Wolter, Nucl. Phys. A732, 24 (2004).

\bibitem{LiQF05b} Q.F. Li, Z.X. Li, S. Soff, R.K. Gupta, M. Bleicher and H.
St\"{o}cker, Phys. Rev. C72, 034613 (2005); ibid, nucl-th/0509070.

\bibitem{Fer05} G. Ferini, M. Colonna, T. Gaitanos, M. Di Toro,
nucl-th/0504032, Nucl. Phys. A (2005) in press.

\bibitem{LiBA05e} Bao-An Li, Lie-Wen Chen, Champak B. Das, Subal Das Gupta,
Charles Gale, Che Ming Ko, Gao-Chan Yong, Wei Zuo, Proc. AIP 791, 22 (2005);
arXiv:nucl-th/0504069.

\bibitem{Yon07} G.C. Yong, B.A. Li, and L.W. Chen, arXiv:nucl-th/0703042.

\bibitem{Ber88b} G.F. Bertsch and S. Das Gupta, Phys. Rep. \textbf{160}
(1988) 189.

\bibitem{Aic91} J. Aichelin, Phys. Rep. \textbf{202} (1991) 233.

\bibitem{LiBA05c} B.A. Li and L.W. Chen, Phys. Rev. C72, 064611 (2005).

\bibitem{Tsa04} M.B. Tsang et al., Phys. Rev. Lett. 92, 062701 (2004).

\bibitem{Che05a} L.W. Chen, C.M. Ko, and B.A. Li, Phys. Rev. Lett. 94,
032701 (2005).

\bibitem{Ste05b} A.W. Steiner and B.A. Li, Phys. Rev. C72, 041601 (R) (2005).

\bibitem{LiBA06a} B.A. Li and Andrew W. Steiner, Phys. Lett. B642 (2006) 436.

\bibitem{Che05b} L.W. Chen, C.M. Ko and B.A. Li, Phys. Rev. C 72 (2005)
064309

\bibitem{Tod05} B.G. Todd-Rutel and J. Piekarewicz,, Phys. Rev. Lett. 95,
122501 (2005).

\bibitem{She07} D. Shetty, S.J. Yennello and G.A. Souliotis, Phys. Rev. C 75
(2007) 034602.

\bibitem{Pra85} M. Prakash and K. S. Bedell, Phys. Rev. C 32, 1118 (1985).

\bibitem{Bar85} E. Baron, J. Cooperstein and S. Kahana, Phys. Rev. Lett.
\textbf{55}, 126 (1985); Nucl. Phys. \textbf{A440}, 744 (1985).

\bibitem{Kah89} S.H. Kahana, Ann. Rev. Nucl. Part. Sci., \textbf{39}, 231
(1989).

\bibitem{Lat91} J.M. Lattimer, C.J. Pethick, M. Prakash and P. Haensel,
Phys. Rev. Lett. \textbf{66}, 2701 (1991).

\bibitem{Sum94} K. Sumiyoshi and H. Toki, Astro. Phys. Journal, \textbf{422},
700 (1994).

\bibitem{Lee96} C-H. Lee, Phys. Rep. \textbf{275}, 255 (1996).

\bibitem{Das03} C.B. Das, S. Das Gupta, C. Gale, and B.A. Li, Phys. Rev. C
67, 034611 (2003).

\bibitem{Wir88b} R.B. Wiringa, Phys. Rev. C38, 2967 (1988).

\bibitem{Zuo05} W. Zuo, L.G. Gao, B.A. Li, U. Lombardo and C.W. Shen, Phys.
Rev. C72, 014005 (2005).

\bibitem{Sat69} G.R. Satchler, Chapter 9: Isospin Dependence of Optical
Model Potentials, in Isospin in Nuclear Physics, page 391-456, D.H.
Wilkinson (Ed.), (North-Holland, Amsterdam, 1969).

\bibitem{Hof72} G.W. Hoffmann and W.R. Coker, Phys. Rev. Lett. 29, 227
(1972).

\bibitem{Hod94} P.E. Hodgson, The Nucleon Optical Model, pages 613-651,
(World Scientific, Singapore, 1994).

\bibitem{Kon03} A.J. Koning and J.P. Delarocje, Nucl. Phys. A713, 231 (2003).

\bibitem{Jam89} M. Jaminon and C. Mahaux, Phys. Rev. C40, 354 (1989).

\bibitem{Neg98} J.W. Negele and H. Orland, Quantum Many-Particle System,
Perseus Books Publishing, L.L.C., 1998

\bibitem{LiBA04c} B.A. Li, Phys. Rev. C69, 064602 (2004).

\bibitem{Beh05} B. Behera, T.R. Routray, A. Pradhan, S.K. Patra and P.K.
Sahu, Nucl., Phys. A753, 367 (2005).

\bibitem{Sjo76} O. Sj\"{o}berg, Nucl. Phys. A265, 511 (1976).

\bibitem{Ber84} G.F. Bertsch, H. Kruse and S. Das Gupta, Phys. Rev. C29, 673
(1984).

\bibitem{Neg81} J.W. Negele and K. Yazaki, Phys. Rev. Lett. 62, 71 (1981).

\bibitem{Pan91} V.R. Pandharipande and S.C. Pieper, Phys. Rev. C45, 791
(1991).

\bibitem{Li93} G.Q. Li and R. Machleidt, Phys. Rev. C48, 1702 (1993); ibid,
C49, 566 (1994).

\bibitem{Sch97} H.-J. Schulze et al., Phys. Rev. C55, 3006 (1997); A.
Schnell et al., ibid, C57, 806 (1998).

\bibitem{Per02} D. Persram and C. Gale, Phys. Rev. C65, 064611 (2002).

\bibitem{Gia96} G. Giansiracusa, U. Lombardo, and N. Sandulescu, Phys. Rev.
C53, R1478 (1996).

\bibitem{Koh98} M. Kohno, M. Higashi, Y. Watanabe, and M. Kawai, Phys. Rev.
C57, 3495 (1998).

\bibitem{LiQF00} Qingfeng Li, Zhuxia Li, and Guangjun Mao, Phys. Rev. C62,
014606 (2000).

\bibitem{Che01} L.W. Chen et al., Phys. Rev. C64, 064315 (2001).

\bibitem{Dan02b} P. Danielewicz, Acta. Phys. Polon. B33, 45 (2002) and
references therein.

\bibitem{Sam05b} F. Sammrruca and P. Krastev, nucl-th/0506081.

\bibitem{Ber88a} G.F. Bertsch, G.E. Brown, V. Koch and B.A. Li, Nucl. phys.
A490, 745 (1988).

\bibitem{Mao94} G.J. Mao, Z.X. Li, Y.Z. Zhuo, Y.L. Han and Z.Q. Yu, Phys.
Rev. C49, 3137 (1994); G.G. Mao, Z.X. Li and Y.Z. Zhuo, ibid, C53, 2933
(1996); C55, 792 (1997).

\bibitem{Gai05} T. Gaitanos, C. Fuchs and H.H. Wolter, Phys. Lett, B609, 241
(2005).

\bibitem{Pra88} M. Prakash, T.T. S. Kuo and S. Das Gupta, Phys. Rev. C37,
2253 (1988).

\bibitem{Far91} M. Farine et al., Z. Phys. A339, 363 (1991).

\bibitem{Gal87} C. Gale, G. Bertsch, and S. Das Gupta, Phys. Rev. C 35, 1666
(1987).

\bibitem{Wel88} G.M. Welke et al.,
Phys. Rev., C 38, 2101 (1988).

\bibitem{Gal90} C. Gale et al., 
Phys. Rev., C 41, 1545 (1990).

\bibitem{Pan93} Q. Pan and P. Danielewicz, Phys. Rev. Lett. 70, 2062 (1993).

\bibitem{Zha94} J. Zhang et al., 
Phys. Rev. C 50, 1617 (1994).

\bibitem{Gre99} V. Greco et al., 
Phys. Rev. C 59, 810 (1999).

\bibitem{Dan00} P. Danielewicz, Nucl. Phys. A673, 375 (2000).

\bibitem{LiBA04d} B.A. Li, Phys. Rev. C 69, 064602 (2004).

\bibitem{Ram00} F. Rami et al., Phys. Rev. Lett. 84, 1120 (2000).

\bibitem{Fuj05} M. Fujiwara, private communications, 2005.

\bibitem{Pie05} J. Piekarewicz, private communications, 2005.

\bibitem{Col05} G. Colo, private communications, 2005.

\bibitem{LiBA05d} B.A. Li, P. Danielewicz, and W.G. Lynch, Phys. Rev. C 71,
054603 (2005).

\bibitem{Bro00} B.A. Brown, Phys. Rev. Lett. 85, 5296 (2000).

\bibitem{Hor01a} C.J. Horowitz, and J. Piekarewicz, Phys. Rev. Lett 86, 5647
(2001); Phys. Rev. C 66, 055803 (2002).

\bibitem{Typ01} S. Typel and B.A. Brown, Phys. Rev. C 64, 027302 (2001).

\bibitem{Fur02} R.J. Furnstahl, Nucl. Phys. A706, 85 (2002).

\bibitem{Kar02} S. Karataglidis, K. Amos, B.A. Brown, and P.K. Deb, Phys.
Rev. C 65, 044306 (2002).

\bibitem{Fri86} J. Friedrich and P.-G. Reinhard, Phys. Rev. C 33, 335 (1986).

\bibitem{Bro98} B.A. Brown, Phys. Rev. C 58, 220 (1998).

\bibitem{Che99b} L.W. Chen and F.S. Zhang, High Energy Phys. and Nucl. Phys.
23, 1197 (1999) (in Chinese).

\bibitem{Sto03} J.R. Stone, J.C. Miller, R. Koncewicz, P.D. Stevenson and
M.R. Strayer, Phys. Rev. C 68, 034324 (2003).

\bibitem{Sta94} V.E. Starodubsky and N.M. Hintz, Phys. Rev. C 49, 2118
(1994).

\bibitem{Cla03} B.C. Clark, L.J. Kerr, and S. Hama, Phys. Rev. C 67, 054605
(2003).

\bibitem{Kra04} \ A. Krasznahorkay et al., Nucl. Phys. A731, 224 (2004).

\bibitem{Hor01b} C.J. Horowitz, and J. Piekarewicz, Phys Rev. C 63, 025501
(2001).

\bibitem{Jef00} Jefferson Laboratory Experiment E-00-003, spokesperson R.
Michaels, P.A. Souder, and G.M. Urciuoli.

\bibitem{Yak06} K. Yako, H. Sagawa and H. Sakai, Phys Rev. C 74, 051303(R)
(2006).

\bibitem{LiBA03} B.A. Li, Phys. Rev. C67, 017601 (2003).

\bibitem{Ber80} G.F. Bertsch, Nature 283, 280 (1980); A. Bonasera and G.F.
Bertsch, Phys. Let. B195, 521 (1987).

\bibitem{Lyn06} W.G. Lynch, L.G. Sobotka and M.B. Tsang, private
communications, (2006); M. A. Famiano et al., Phys. Rev. Lett. 97,
052701 (2006) .

\bibitem{Sto86a} R. Stock, Phys. Rep. 135, 259 (1986).

\bibitem{Li91} B. A. Li and W. Bauer, Phys. Rev. C44, 450 (1991).

\bibitem{LiBA06b} B.A. Li, L.W. Chen, G.C. Yong, and W. Zuo, Phys. Lett.
B634, 378 (2006).

\bibitem{Yon06a} G.C. Yong, B.A. Li, L.W. Chen, and W. Zuo, Phys. Rev. C 73,
034603 (2006).

\bibitem{Sto86b} H. St\"ocker and W. Greiner, Phys. Rep. \textbf{137} (1986)
277.

\bibitem{Cas90} W. Cassing, V. Metag, U. Mosel and K. Niita,Phys. Rep.
\textbf{188} (1990) 363.

\bibitem{Rei97} W. Reisdorf and H. G. Ritter, Annu. Rev. Nucl. Part. Sci.
\textbf{47} (1997) 663.

\bibitem{Yon06b} G.C. Yong, B.A. Li, L.W. Chen, Phys. Rev. C \textbf{74}
(2006) 064617.

\bibitem{Htu99} M.M. Htun \textsl{et al}., Phys. Rev. C \textbf{59} (1999)
336 and references therein.

\bibitem{Ven93} L. Venema \textsl{et al}., Phys. Rev. Lett. \textbf{71}
(1993) 835.

\bibitem{Lei93} Y. Leifels \textsl{et al}., Phys. Rev. Lett. \textbf{71}
(1993) 963.

\bibitem{Lam93} D. Lambrecht \textsl{et al}., Z. Phys. \textbf{A350} (1994)
115.

\bibitem{Bas95} S.A. Bass, C. Hartnack, H. St\"ocker and W. Greiner, Z.
Phys. \textbf{A352} (1995) 171.

\bibitem{Lar00} A. B. Larionov, W. Cassing, C. Greiner and U. Mosel, Phys.
Rev. C \textbf{62} (2000) 064611.

\bibitem{Fam06} M.A. Famiano \textsl{et al}., Phys. Rev. Lett. \textbf{97}
(2006) 052701.

\bibitem{Bic07} A. Bickley, M.A. Famiano, W.G. Lynch, G.D. Westfall
\textsl{et al}., private communications and their talks at the 2007
Town Meeting for the NSAC Long Range Plan, Chicago, January 19-21,
2007.

\bibitem{Aic85} J. Aichelin and C.M. Ko, Phys. Rev. Lett. \textbf{55} (1985)
2661.

\bibitem{Ko96} C.M. Ko and G.Q. Li, J. Phys. G \textbf{22}, 1673 (1996).

\bibitem{Ko97} C.M. Ko, V. Koch and G.Q. Li, Ann. Rev. Nucl. Part. Sci.
\textbf{47}, 505 (1997).

\bibitem{Cas99} W. Cassing and E. L. Bratkovskaya, Phys. Rep. \textbf{308}
(1999) 65.

\bibitem{Kol05} E.E. Kolomeitsev, C. Hartnack, H.W. Barz, M. Bleicher, E.
Bratkovskaya, W. Cassing, L.W. Chen, P. Danielewicz, C. Fuchs, T. Gaitanos,
C.M. Ko, A. Larionov, M. Reiter, Gy. Wolf, J. Aichelin, J. Phys. G31 (2005)
S741-S758.

\bibitem{Fuc06a} C. Fuchs, Prog. Part. Nucl. Phys. \textbf{56}, (2006) 1

\bibitem{LiQF05c} Q. Li, Z. X. Li, S. Soff, Raj K. Gupta, M. Bleicher and H.
Stoecker, J. Phys. G 31 (2005) 1359-1374.

\bibitem{Lop07} X. Lopez, Y.J. Kim, N. Herrmann et al. (FOPI Collaboration),
Phys. Rev. C \textbf{75} (2007) 011901(R).

\bibitem{Fuc06b} C. Fuchs, H.H. Wolter, Eur. Phys. J. A \textbf{30} (2006) 5.
\end{thebibliography}
\end{document}